\documentclass[12pt]{article}
\usepackage{amsmath, amssymb, slashed,  color}
\usepackage{graphicx}
\usepackage{stmaryrd}

\def\a{\alpha}
\def\b{\beta}

\def\d{\delta}

\def\g{\gamma}

\def\k{\kappa}
\def\l{\lambda}
\def\m{\mu}
\def\n{\nu}
\def\p{\psi}

\def\r{\rho}
\def\s{\sigma}

\def\w{\omega}

\def\z{\zeta}
\def\D{\Delta}
\def\G{\Gamma}

\def\L{\Lambda}
\def\F{\Phi}

\def\W{\Omega}

\def\del{\partial}
\def\LRarrow{\leftrightarrow}

\def\go{\overset{{}_{\shortrightarrow}}}
\def\back{\overset{{}_{\shortleftarrow}}}

\newcommand{\1}{\mbox{1}\hspace{-0.25em}\mbox{l}}

\begin{document}
%%%%%%%%%%%%%%%%%%%%%%%%%%%%%%%%%%
%%%%%%%%%%% Title page %%%%%%%%%%%
%%%%%%%%%%%%%%%%%%%%%%%%%%%%%%%%%%
\begin{titlepage}
\begin{center}

\hfill ICRR-report-700-2015-1\\
\hfill IPMU15-0048 \
%\hfill \today

\vspace{2.0cm}
{\large\bf Spontaneous thermal Leptogenesis via Majoron oscillation}

\vspace{2.0cm}
{\bf Masahiro Ibe}$^{a,b}$ and
{\bf Kunio Kaneta}$^{a}$

\vspace{1.5cm}
{\it
$^{(a)}${ICRR, the University of Tokyo, Kashiwa, Chiba 277-8582, Japan}
}\\
{\it
$^{(b)}${Kavli IPMU (WPI), TODIAS, the University of Tokyo, Kashiwa, Chiba 277-8583, Japan}
}

\vspace{2.5cm}
\abstract{
A novel model of spontaneous Leptogenesis is investigated, where it takes place 
in the thermal equilibrium due to a background Nambu-Goldstone field in motion.
In particular, we identify the Nambu-Goldstone field to be the Majoron 
which associates with spontaneous breakdown of (discrete) $B-L$ symmetry.
In this scenario sufficient lepton number asymmetry is generated in primordial thermal bath 
without having $CP$-violating out-of-equilibrium decay of the heavy right-handed Majorana 
neutrinos.
To obtain the observed baryon asymmetry, the neutrino masses are predicted in certain ranges, 
which can be translated into the effective mass of the neutrinoless double beta decay.
}

\end{center}
\end{titlepage}
\setcounter{footnote}{0}

%%%%%%%%%%%%%%%%%%%%%%%%%%%%%%%%%%%%
%%%%%%%%%%% Introduction %%%%%%%%%%%
%%%%%%%%%%%%%%%%%%%%%%%%%%%%%%%%%%%%
\section{Introduction}

Baryon asymmetry of the universe  is one of the biggest mysteries in particle physics and cosmology.
Direct observations show that the universe contains no appreciable primordial anti-matter, although
theories of particle physics treat matter and anti-matter in an equitable manner~\cite{Cohen:1997ac}.
To date, the asymmetry between matter and anti-matter has been  measured precisely by two independent observations, 
the cosmic microwave background (CMB) measurement~\cite{Planck:2015xua} 
and the measurements of the primordial abundances of the light elements~\cite{Cooke:2013cba} predicted by Big Bang Nucleosynthesis (BBN).
The baryon asymmetry is often parametrized by the baryon abundance, $\W_Bh^2$,
\begin{eqnarray}
	\W_Bh^2 = 0.0222^{+0.00045}_{-0.00043}~~~(Planck~\cite{Planck:2015xua}),
	~~~0.0220\pm0.046~~~(\text{BBN}~\cite{Cooke:2013cba})\ ,
\end{eqnarray}
or parametrized by the baryon-to-photon number ratio at today's temperature of the universe, $\eta_B$, 
\begin{eqnarray}
\eta_B \simeq  6\times 10^{-10}\times \left(\frac{\W_Bh^2}{0.022}\right)\ .
\end{eqnarray}

In view of the success of the inflationary paradigm, 
the baryon asymmetry generated before inflation has been diluted away,
and hence, the baryon asymmetry cannot be explained by an initial condition of the universe.
As pointed out by Sakharov~\cite{Sakharov:1967dj}, to generate the asymmetry dynamically, following three conditions 
are required to be fulfilled in the expanding universe; (i) baryon number non-conservation ($B$ violation), (ii) $C$ and $CP$ violation, (iii) departure form thermal equilibrium.
The standard model (SM) could satisfy the all of the conditions, where (i) is achieved by quantum effect known as {\it sphaleron}~\cite{Klinkhamer:1984di}, (ii) is provided by the $CP$ violating phase in the Cabbibo-Kobayashi-Masukawa (CKM) matrix, and the first order electroweak phase transition could be realized (iii) if the Higgs boson has an appropriate mass.
However, the SM fails to generate sufficient baryon asymmetry since the $CP$ violating phase in the CKM matrix is too small, and the observed Higgs mass 125 GeV is too heavy for (iii)~\cite{Shaposhnikov:1987tw}.
It is therefore of importance to explore the origin of the baryon asymmetry in physics beyond the SM.

Among various possibilities of Baryogengesis associated with  physics beyond the SM,
Leptogenesis is one of the most attractive mechanisms~\cite{Fukugita1986}.
There, $B$ violation is provided by  lepton ($L$) symmetry breaking of the right-handed neutrino mass
in conjunction with the  {\it sphaleron} effect, 
and the condition (ii), the $CP$ violation, is satisfied by the $CP$-violating phases of the neutrino Yukawa couplings.
Then, lepton number asymmetry is generated when the right-handed neutrinos in the thermal bath 
decay slowly in an out-of-equilibrium way so that the condition (iii) is satisfied. 
Leptogenesis is quite attractive since it is naturally achieved along with the seesaw mechanism~\cite{Yanagida:1900hni,GellMann:1980vs,Glashow:1979,Mohapatra:1979ia,Schechter:1980gr} (see also~\cite{Minkowski:1977sc}) which explains the tiny neutrino masses 
by the heaviness of the right-handed Majorana neutrinos.

In this paper, we investigate an alternative model of Leptogenesis which does not rely 
on the out-of-equilibrium decay of the right-handed neutrinos.
Instead, we consider a model of spontaneous Baryogengesis which has been originally proposed  by Cohen and Kaplan in 
Ref.~\cite{Cohen1987,Cohen1988}, where, in our case, Leptogenesis takes place 
in the thermal equilibrium due to a background Majoron field in motion.
In general spontaneous Baryogenesis scenario, the background field in motion causes
level splittings between the matter and the anti-matter 
via its derivative couplings to the current of  $B$ symmetry.%
\footnote{In the literature, the level splitting is often called the effective chemical potential.
However, this terminology is somewhat misleading, since the chemical potential is defined to characterize the thermal bath.
In this paper, we use ``(dynamical) level splitting" to refer the effective chemical potential appearing in spontaneous Baryogenesis.}
Then, with efficient baryon violating processes in the thermal equilibrium, the non-vanishing level splitting 
leads to the baryon asymmetry of the universe.
This mechanism for generating baryon asymmetry works in various setups~\cite{Yamaguchi2002,Chiba:2003vp,Davoudiasl2004}, and recently it is discussed that the lepton asymmetry, i.e. spontaneous Leptogenesis, can be also achived via anomalous symmetries~\cite{Kusenko:2014lra,Kusenko:2014uta}.

A notable feature of our model is the use of the Majoron which is the Nambu-Goldstone field
associated with spontaneous (discrete) $B-L$ symmetry breaking~\cite{Chikashige1980,Gelmini:1980re}.%
\footnote{See, for example, ~\cite{Schechter:1981cv,Dearborn:1985gp,Kim:1986ax,Chang:1988aa,Choi:1989hi,Berezinsky:1993fm,Choudhury:1993hv,Queiroz:2014yna,Sierra:2014sta,Boucenna:2014uma}
for extensive studies on phenomenological and cosmological aspects of light Majoron.}
By remembering that spontaneous $B-L$ breaking is inevitable to obtain the mass of the heavy right-handed neutrinos,
this choice makes the model more economical than the models with additionally introduced field.
Moreover, it is also advantageous that the derivative couplings of the Majoron to the SM fields 
are an automatic consequence of the spontaneous $B-L$ breaking where 
they are controlled by the $B-L$ charges of the SM fields.

This paper is organized as follows.
In section 2, we explain our setup, and gives a formulation to make our notation clear.
Our main results are given in section 3 where numerical analysis of the baryon abundance and some phenomenological implications are discussed.
Section 4 devotes a summary.

\section{Spontaneous $B-L$ symmetry breaking and Majoron oscillation}
\label{sec:SSB and Majoron oscillation}
In this section we specify the model of our interest, in which right-handed neutrinos $N_R$ 
acquire  super heavy Majorana masses via spontaneous (discrete) $B-L$ symmetry breaking.
For now, let us confine ourselves to a minimal extension of the SM with  global $U(1)_{B-L}$ symmetry 
and one flavor approximation for the lepton sector.
To achieve the Majorana neutrino mass from spontaneous breaking  of $B-L$ symmetry, 
we introduce a SM singlet complex scalar field, $\s$, with the $B-L$ charge $-2$. 
With this charge assignment, the Lagrangian is given by,
\begin{eqnarray}
	{\cal L} 
	&=& 
	{\cal L}_{SM} 
	+ \frac{1}{2}\bar N_R i\slashed{\del}N_R
	+ |\del_\m\s|^2
	- \left[y_\n \bar N_R (L\cdot H) + h.c.\right]\nonumber\\
	& &
	- \frac{g_N}{2}\left[\s \overline{N_R^C} N_R + h.c.\right]
	- V(H,\s),\\
	V(H,\s)
	&=&
	\l_H (|H|^2 - v_{ew}^2)^2 + \l_\s (|\s|^2 - v_{B-L}^2)^2 \nonumber\\
	& & + \l_{\s H} (|\s|^2 - v_{B-L}^2)(|H|^2 - v_{ew}^2)\ . 
\end{eqnarray}
Here,  $y_\nu$, $g_N$, $\lambda_{H,\s,\s H}$ are dimensionless coupling constants,
while $v_{ew}$ and $v_{B-L}$ are dimensionful parameters which provide the
electroweak and the $B-L$ symmetry breaking scales, respectively.
The dot products denote the $SU(2)$ invariant products, $(A\cdot B)\equiv A_1B_2-A_2B_1$.
Around the ${B-L}$ breaking vacuum, we parametrize $\s$ by
\begin{eqnarray}
	\s &=& (v_{B-L} + \r/\sqrt{2})e^{i\chi/(\sqrt{2}v_{B-L})},
\end{eqnarray}
where $\r$ is a real part of $\s$, and $\chi$ corresponds to the Nambu-Goldstone boson, i.e. the Majoron.
After $B-L$ symmetry breaking, $N_R$ acquires the Majorana mass $M_R=g_N v_{B-L}$. 
Hereafter, we assume $g_N = O(1)$ and omit  $g_N$, i.e., $v_{B-L} = M_R$. 

Since we are mainly interested in the universe where the right-handed neutrinos have decoupled from the thermal bath, 
it is enough to use an effective theory obtained by integrating out $N_R$:
\begin{eqnarray}
	{\cal L}_{\rm eff}
	&=&
	\text{(kinetic terms)} 
	- \frac{\del_\m\chi}{\sqrt{2}M_R}J^\m_L - \frac{m_\n}{2v_{ew}^2}(\overline{L^C}\cdot H)(L\cdot H) + \cdots, \label{Leff}
	\label{eq:dim5}
\end{eqnarray}
to take the $L$ breaking effects into account.
Here, $J^\m_L$ is the fermionic lepton current, and $m_\n=|y_\n|^2v_{ew}^2/M_R$ denotes the mass of the light neutrinos
in the SM.
To achieve this Lagrangian, we have redefined $N_R\to e^{i\chi/(2\sqrt 2v_{B-L})}N_R$ and $L\to e^{i\chi/(2\sqrt 2v_{B-L})}L$.
It should be noted that these redefinitions are just a choice of basis in field space, and hence, 
physical observables do not depend on the choices of the basis.%
\footnote{It is also possible to perform computations without these redefinitions. 
In this case the current term does not appear, and instead, the Majoron field appears 
in front of  the dimension five operators, $\frac{m_\n}{2v_{ew}^2}e^{-i\chi/(\sqrt{2}v_{B-L})}(\overline{L^C}\cdot H)(L\cdot H)$. See the appendix~\ref{sec:appendix2} for details.}.
As we will see in Section \ref{sec:result}, the temperature $T_{osc}$ at which the Majoron starts to oscillate
and spontaneous Leptogenesis takes place successfully is $T_{osc}\gtrsim 10^{13}$\,GeV (see Eq.~(\ref{eq:Tosc and TBL})).
In the following analysis, we assume that the right-handed neutrinos are heavy enough,
$M_R \gtrsim T_{\rm osc}\gtrsim 10^{13}$\,GeV, so that the effective field theory in Eq.\,(\ref{eq:dim5})
is applicable.

At this stage, Majoron is a massless boson because so far it has been treated as an exact Nambu-Goldstone boson.
It is known, however, that global symmetries might not be respected in context of quantum gravity, and hence violated by gravitational effects~\cite{Giddings:1987cg,Abbott:1989jw,Coleman:1989zu}.
Thus, once gravitational effect turns on, Majoron would acquires a mass~\cite{Akhmedov:1992hi,Rothstein:1992rh} via Planck scale suppressed operators:
\begin{eqnarray}
	{\cal O}_M^{(n)} &=& \frac{\s^n}{M_{\rm Pl}^{n-4}}~~~(n=5,6,7,\cdots),
	\label{Omass}
\end{eqnarray}
where $M_{\rm Pl}\simeq 2.4\times 10^{18}$ GeV is the reduced Planck mass.
We assume that these breaking operators appear with $O(1)$ coefficients which we have omitted here.
It turns out that the Majoron mass induced by the dimension $n$ operator, $m_\chi^{(n)}$, is given by
\begin{eqnarray}
	m_\chi^{(n)} &\sim& \left(\frac{M_R^{n-2}}{M_{\rm Pl}^{n-4}}\right)^{1/2}.
	\label{mchi}
\end{eqnarray}
It should be  commented that the origin of the explicit breaking terms in Eq.\,(\ref{Omass}) 
may be also attributed to  spontaneous breaking of the gauged and hence $U(1)_{B-L}$ symmetry 
broken at around the Planck scale to a discrete $B-L$ ($Z_{2n}$) symmetry.
In the following section, we discuss how the motion of Majoron leads to spontaneous Leptogenesis 
in this setup. 

For preparation, let us scketch how the Majoron field behaves in the expanding universe.
We suppose that $B-L$ symmetry is spontaneously broken during/before inflation.
During inflation, the quantum fluctuation of the Majoron is exponentially stretched~\cite{Bunch:1978yq,Linde:1982uu,Starobinsky:1994bd}, and the Majoron field is settled at  some point on its field space.
After inflation, the Majoron field behaves according to the equation of motion,
\begin{eqnarray}
	\ddot\chi + 3H \dot\chi = -\del_\chi V_{\rm eff},\label{EOM}
\end{eqnarray}
where $H$ is the Hubble parameter. Here we define $\del_0\chi\equiv \dot\chi$, and take $V_{\rm eff}\simeq (m_\chi^2/2)\chi^2$ with the Majoron mass $m_\chi$.
In the radiation dominant epoch, the solution of Eq.~(\ref{EOM}) is obtained as
\begin{eqnarray}
	\chi(t) &=&\sqrt{2} \chi_0\, \G\left(\frac{5}{4}\right)\left(\frac{2}{m_\chi t}\right)^{1/4} J_{1/4}(m_\chi t)\ ,
\end{eqnarray}
where, $\sqrt{2}\chi_0$ is the initial amplitude.
This shows that the Majoron starts to oscillate coherently when the Hubble parameter decreases down to $H\sim m_{\chi}$.
During inflation, the Majoron is expected to take a field value on its field space, $0 < \chi/\sqrt{2} \leq \pi v_{B-L}$,
and hence, the initial value of the motion of the Majoron is $\chi_0=O(M_R)$.

Once the Majoron starts to oscillate, the derivative couplings in Eq.~(\ref{Leff}) become $\dot\chi/(\sqrt{2}M_R)J^0_L$, 
which eventually lead to level splittings between the leptons and the anti-leptons as we will discuss shortly.
A notable feature of our setup is that the Lepton number violating processes automatically present in Eq.\,(\ref{eq:dim5})
as the dimension five neutrino mass operators, and hence, all the necessary ingredients for spontaneous Leptogenesis,
i.e. derivative couplings to the Majoron and the lepton number violating interactions, are equipped with. 
As we will see in the next section, once we specify the neutrino masses, resultant lepton asymmetry depends on only two parameters, i.e., the mass and the initial amplitude of the Majoron.
In particular, certain value of the initial amplitude is automatically chosen due to inflation since $B-L$ symmetry breaking occurs before/during inflation in our setup.
Although the initial value is randomly determined, it is typically to be of the order of $B-L$ violation scale, $O(M_R)$.
It should be also emphasized that the dynamical level splittings violate the $CPT$-invariance, 
and hence, baryon asymmetry can be generated without satisfying the Sakharov's conditions exactly.

\section{Spontaneous thermal Leptogenesis}
As we have seen in the previous section, the motion of the background Majoron field leads to non-trivial
contributions to the kinetic term of the leptons.
In this section, we first discuss the kinematics of leptons in the presence of the background Majoron field. 
Then, we will move on to explore viable parameter regions by solving the Boltzmann equation for the lepton asymmetry.

\subsection{Thermally averaged cross sections in the presence of  Majoron background in motion}
To see how the level splittings appear, let us look at the kinetic terms of the leptons.
Here, let us collectively denote the charged leptons and the neutrinos by $\p$ with a mass $m$.%
\footnote{Here, we discuss the case of the Dirac fermions, although the result can be applied 
to the Majorana fermions.}
Due to the derivative coupling to the Majoron, the kinetic term of $\psi$ is deformed
by $\m_\chi\equiv \dot{\chi}/(\sqrt{2}M_R)$:
\begin{eqnarray}
	{\cal L}_{\rm kin} &=& \bar \p (i\slashed{\del}-m-\m_\chi\g^0)\p\ .
	\label{Lkin}
\end{eqnarray}
For $\p$ having momentum $p=(E,\vec p)$, the dispersion relation of $\p$ changes to $E=\pm\sqrt{m^2+|\vec p|^2}+\m_\chi$ due to non-zero contribution of $\m_\chi$, contrary to $E^0=\pm\sqrt{m^2+|\vec p|^2}$ for $\m_\chi=0$.
Thus, the term proportional to $\mu_\chi$ causes the energy level splittings between the leptons and the anti-leptons.%
\footnote{See the appendix \ref{sec:appendix1} for detailed discussion.}

In the presence of the dynamical level splittings, the lepton number asymmetry is generated 
in thermal equilibrium by the $L$ symmetry violating processes
via the dimension five operator in Eq.~(\ref{Leff}).
That is, as we will show by solving the Boltzmann equations, 
 the $L$  breaking processes smooth out imbalances in the lepton and the anti-lepton numbers caused by 
the level splittings. 

By denoting $L=(\n_L,e_L^-)^T$ and $H=(h^+,h^0)^T$ the relevant processes for the Boltzmann equation
of the lepton asymmetry are as follows,
%\footnote{Here, we neglect the effects of electroweak symmetry breaking since we are interesting 
%in the cosmic temperature much higher than than $v_{ew}$.}:
\begin{align}
	&
	(a)~~ h^+h^+ \LRarrow e_L^+e_L^+,~~~
	(b)~~ h^-h^- \LRarrow e_L^-e_L^-,~~~
	(c)~~ h^0h^0 \LRarrow \bar\n_L\bar\n_L,~~~
	(d)~~ h^{0\dagger}h^{0\dagger} \LRarrow \n_L\n_L,\nonumber\\
	&
	(e)~~ h^+h^0 \LRarrow e_L^+\bar\n_L,~~~
	(f)~~ h^-h^{0\dagger} \LRarrow e_L^-\n_L,\nonumber\\
%------------------------------------------------------------------
	&
	(g)~~ h^+e_L^- \LRarrow h^-e_L^+,~~~
	(h)~~ h^0\n_L \LRarrow h^{0\dagger}\bar\n_L,~~~
	(i)~~ h^+e_L^- \LRarrow h^{0\dagger}\bar\n_L,~~~
	(j)~~ h^-e_L^+ \LRarrow h^0\n_L,\nonumber\\
	&
	(k)~~ h^+\n_L \LRarrow h^{0\dagger}e_L^+,~~~
	(l)~~ h^-\bar\n_L \LRarrow h^0e_L^-.\nonumber
\end{align}
We express the scattering amplitudes for the process, for example, $h^+h^+\to e_L^+e_L^+$ by ${\cal M}_{\go a}$, and $e_L^+e_L^+\to h^+h^+$ by ${\cal M}_{\back a}$.
To obtain the Boltzmann equations, it is convenient to consider thermally averaged cross sections in which we approximate 
the distribution functions by the Maxwell-Boltzmann distribution,
\begin{eqnarray}
	\langle\s_I v\rangle 
	&\equiv&
	(n^{\rm eq})^{-2}\int d\Pi_1d\Pi_2d\Pi_3d\Pi_4e^{-E^0_{1}/T}e^{-E^0_{2}/T}\sum_{\rm spins}|{\cal M}_I|^2,
\end{eqnarray}
where $I=\go a, \back a, \cdots$, and we define $d\Pi_i=d^3p_i/(2\pi)^3/(2E^{0}_i)$ for the momenta assigned by 
$p_1p_2\to p_3 p_4$ in each process.%
\footnote{Here, we define the chemical potential of the thermal equillibrium
so that the Boltzmann factors are given by $e^{-(E_{1,2}-\mu_\psi)/T} = e^{-E^0_{1,2}/T}$.
See the appendix A for details.}
Normalization factor $n^{\rm eq}$ is given by $n^{\rm eq}=gT^3/\pi^2$ where $g$ is degrees of freedom of corresponding particle.

The detailed discussion of the cross sections are shown in the appendix~\ref{sec:appendix1}, and finally we obtain $\langle\s_I v\rangle$ as follows:
\begin{eqnarray}
	&
	\langle\s_{\vec a} v\rangle
	\simeq
	\langle\s_0 v\rangle
\displaystyle\left[
	1 + \frac{1}{2}\frac{\m_\chi}{T}
	\right],~~~
	\langle\s_{\back a} v\rangle
	\simeq
	\langle\s_0 v\rangle,
		\hspace{1.8cm}\\
	&
	\langle\s_{\go g} v\rangle
	\simeq
	\langle\s_0 v\rangle
\displaystyle	\left[
	1 - \frac{5}{4}\frac{\m_\chi}{T}
	\right],~~~
	\langle\s_{\back g} v\rangle
	\simeq
	\langle\s_0 v\rangle
\displaystyle	\left[
	1 + \frac{5}{4}\frac{\m_\chi}{T}
	\right],
\end{eqnarray}
where we take massless limit.
Other cross sections can be written in terms of them:
\begin{eqnarray}
	&&
 	\langle\s_{\go a}v\rangle
 	=
 	\langle\s_{\go b}v\rangle
 	=
 	\langle\s_{\go c}v\rangle
 	=
 	\langle\s_{\go d}v\rangle
 	=
 	(1/4)\langle\s_{\go e}v\rangle
 	=
 	(1/4)\langle\s_{\go f}v\rangle,\\
 	&&
 	\langle\s_{\back a}v\rangle
 	=
 	\langle\s_{\back b}v\rangle
 	=
 	\langle\s_{\back c}v\rangle
 	=
 	\langle\s_{\back d}v\rangle
 	=
 	(1/4)\langle\s_{\go e}v\rangle
 	=
 	(1/4)\langle\s_{\go f}v\rangle,\\
 	&&
 	\langle\s_{\go g}v\rangle
 	=
 	\langle\s_{\go h}v\rangle
 	=
 	(1/4)\langle\s_{\go i}v\rangle
 	=
 	(1/4)\langle\s_{\back j}v\rangle
 	=
 	(1/4)\langle\s_{\go k}v\rangle
 	=
 	(1/4)\langle\s_{\back l}v\rangle,\\
 	&&
 	\langle\s_{\back g}v\rangle
 	=
 	\langle\s_{\back h}v\rangle
 	=
 	(1/4)\langle\s_{\back i}v\rangle
 	=
 	(1/4)\langle\s_{\go j}v\rangle
 	=
 	(1/4)\langle\s_{\back k}v\rangle
 	=
 	(1/4)\langle\s_{\go l}v\rangle,
 \end{eqnarray} 
where $\langle\s_0v\rangle\simeq m_\n^2/(32\pi v_{ew}^4)$.

\subsection{Boltzmann equations}
Before deriving the Boltzmann equations, let us discuss the relations among the chemical potentials
so that the Boltzmann equations are reduced.
First, let us list the chemical potentials of the SM particles:
\begin{eqnarray}
	\text{gauge bosons}
	&:&
	\m_\g,~~~\m_{W^\pm},~~~\m_Z,~~~\m_g,\nonumber\\
	\text{matter fermions}
	&:&
	\m_{e_{Li}},~~~\m_{\bar e_{Li}},~~~\m_{\n_{Li}},~~~\m_{\bar \n_{Li}},\nonumber\\
	& &
	\m_{e_{Ri}},~~~\m_{\bar e_{Ri}},\nonumber\\
	& &
	\m_{u_{Li}},~~~\m_{\bar u_{Li}},~~~\m_{d_{Li}},~~~\m_{\bar d_{Li}},\nonumber\\
	& &
	\m_{u_{Ri}},~~~\m_{\bar u_{Ri}},~~~\m_{d_{Ri}},~~~\m_{\bar d_{Ri}},\nonumber\\
	\text{Higgs boson}
	&:&
	\m_{h^0},~~~\m_{h^\pm},\nonumber
\end{eqnarray}
where the index $i$ denotes the flavors.%
\footnote{Here, again we are neglecting electroweak symmetry breaking, and hence, 
$\mu_{\g,Z,W^\pm}$ should be understood as the ones of the $SU(2)\times U(1)_Y$ gauge bosons, strictly
speaking.}
In the highly heated thermal bath, the gauge bosons have vanishing chemical potentials,
and hence, the chemical potentials of the particles and the anit-particles take opposite values
with each other.
Further, we hereafter neglect the flavor mixing for simplicity, so that the chemical potentials do not depend on flavors:
\begin{eqnarray}
	& & \m_{e_{Li}} \equiv \m_{e_L},~~~\m_{e_{Ri}} \equiv \m_{e_R},~~~\m_{\n_{Li}} \equiv \m_{\n_L} ,\nonumber\\
	& & \m_{u_{Li}} \equiv \m_{u_L},~~~\m_{d_{Li}} \equiv \m_{d_L},~~~\m_{u_{Ri}} \equiv \m_{u_R},~~~\m_{d_{Ri}} \equiv \m_{d_R} .
\end{eqnarray}
The vanishing chemical potential $\mu_{W^\pm} = 0 $  leads to further reductions,  $\m_{u_L}-\m_{d_L}=\m_{\n_L}-\m_{e_L}=\m_{h^+}-\m_{h^0}=0$.
For later purpose, we introduce $\m_L$, $\m_Q$, and $\m_H$ which denote
\begin{eqnarray}
\label{eq:LQH}
	& & \m_{e_L} = \m_{\n_L} \equiv \m_L,~~~\m_{u_L} = \m_{d_L} \equiv \m_{Q},~~~\m_{h^0} = \m_{h^+} \equiv \m_H.
\end{eqnarray}

Neutrality of the universe also puts a constraint on the chemical potentials.
In general, the total charge of the universe, which is denoted as $Q_A^{\rm tot}$ for a quantum number $A$, is obtained by $Q_A^{\rm tot}=\sum_i\D n_iQ_{Ai}$ 
where $Q_{Ai}$ is a charge of a particle $i$, and $\D n_i$ is defined by $\D n_i\equiv n_i^\text{particle}-n_i^\text{anti-particle}=2g_iT^3/\pi^2(\m_i/T)$ with $\m_i$ and $g_i$ being the chemical potential of particle $i$ and its degrees of freedom, respectively.
Here we have again approximated the distributions by the Maxwell-Boltzmann distribution.
%\footnote{
The asymmetries between the particle and the anti-particle numbers are
given by $1/6$ and $1/3$ instead of $2/\pi^2$ 
for the Fermi-Dirac and for the Bose-Einstein distributions, respectively.
Therefore, our numerical analyses are expected to be saddled with $O(10)$\% errors due to the Maxwell-Boltzmann
approximation. %}
Thus, the total hypercharge, $Q_Y^{\rm tot}$, is given by 
\begin{eqnarray}
	Q_Y^{\rm tot}
	&=&
	\frac{2T^2}{\pi^2}
	\left[
	N_g \left(\frac{1}{2}(\mu_{u_L}+\m_{d_L}) + 2\m_{u_R} - \m_{d_R} - \frac{1}{2}(\m_{e_L} + \m_{\n_L}) - \m_{e_R}\right) \right.\nonumber\\
	& &
	\left.
	+ \frac{N_h}{2} (\m_{h^+} + \m_{h^0})
	\right],
	\label{eq:neutrality}
\end{eqnarray}
where $N_g$ and $N_h$ are the number of generation and Higgs doublet, respectively.
In the period when the electroweak interaction is in equilibrium, 
the Yukawa interactions are also effective in thermal equilibrium%
\footnote{Strictly speaking, the Yukawa interaction of the electron-type is in thermal equilibrium only
for $T \lesssim 10^{4}$\,GeV.}, 
which leads to
\begin{eqnarray}
\label{eq:Yukawa}
	\m_L - \m_{e_R} - \m_H = 0,~~~
	\m_Q - \m_{u_R} + \m_H = 0,~~~
	\m_Q - \m_{d_R} - \m_H = 0\ ,
\end{eqnarray}
where we have used Eq.\,(\ref{eq:LQH}).
Altogether, the neutrality condition for the total hypercharge gives the relation
\begin{eqnarray}
	Q_Y^{\rm tot}
	&=&
	\frac{2T^2}{\pi^2}[2N_g(\m_Q - \m_L) + (4N_g + N_h)\m_H] =0.
\end{eqnarray}
Besides, we suppose that the sphaleron interaction is in thermal equilibrium during Leptogenesis occurs, which leads to
\begin{eqnarray}
	N_g (3\m_Q + \m_L) &=& 0.
\end{eqnarray}

By solving above relations, we can express all the chemical potentials of the SM particles in terms of  $\m_L$;
\begin{eqnarray}
	& &
	\m_H = \frac{8N_g}{3(4N_g + N_h)}\m_L,~~~
	\m_{e_R} = \frac{4N_g + 3N_h}{3(4N_g + N_h)}\m_L,\\
	& &
	\m_{u_R} = \frac{4N_g - N_h}{3(4N_g + N_h)}\m_L,~~~
	\m_{d_R} = -\frac{12N_g + N_h}{3(4N_g + N_h)}\m_L,~~
	\m_Q = -\frac{1}{3} \m_L.
\end{eqnarray}
In particular, the baryon number $n_B=(2T^2/\pi^2)N_g(2\m_Q+\m_{u_R}+\m_{d_R})$ and the lepton number $n_L=(2T^2/\pi^2)N_g(2\m_L+\m_{e_R})$ are given by,
\begin{eqnarray}
	n_B = \frac{2}{\pi^2}T^2\times \left(-\frac{4}{3}N_g\m_L\right),~~~
	n_L = \frac{2}{\pi^2}T^2\times \frac{28N_g + 9N_h}{3(4N_g + N_h)}N_g\m_L.
\end{eqnarray}
From this, we also obtain a relation between baryon asymmetry and lepton asymmetry,
\begin{eqnarray}
	n_B = -\frac{4(4N_g + N_h)}{28N_g + 9 N_h}n_L,
\end{eqnarray}
where $n_B=-(52/93)n_L$ by substituting $N_g=3$ and $N_h=1$ in the SM.%
\footnote{The ratio of the baryon asymmetry to the lepton asymmetry differs from the ones in the literature since the neutrality condition given in Eq.~(\ref{eq:neutrality}) is slightly different due to the Maxwell-Boltzmann approximation.}

As a result of the above reductions, we are left with only one undetermined chemical potential, $\m_L$, and hence, 
we only need to solve the Boltzmann equation of $\m_L$.
By using the cross sections given in the previous subsection, the Boltzmann equation of $\mu_L$ is given by%
\footnote{It should be noted that the obtained differential equation is the Boltzmann equation for the total lepton charge $Q_L^{\rm tot}$, and thus, we can safely omit all collision terms other than $L$ violating terms: $dQ_L^{\rm tot}/dt \propto \langle\s_0v\rangle$.}
\begin{eqnarray}
	\frac{d}{dT}\frac{\m_L}{T}
	&=&
	w \left(\frac{\m_L}{T} - \a \frac{\m_\chi}{T}\right),\label{Boltzmann1}
\end{eqnarray}
where $w$ is the wash-out factor defined by
\begin{eqnarray}
	w \simeq \frac{\pi^4g_{*s}}{90}\frac{ \k \langle\s_0v\rangle}{sHT}\frac{T^6}{\pi^4},~~~
	 \k \sim 3\times10^{2},
\end{eqnarray}
and $\a$ is a numerical factor given by $\a\sim0.5$.
Here $s$ and $H$ are the entropy density and the Hubble parameter given by
\begin{eqnarray}
	s = \frac{2\pi^2g_{*s}}{45}T^3,~~~H=\sqrt{\frac{\pi^2g_*}{90}}\frac{T^2}{M_{\rm Pl}},
\end{eqnarray}
where $g_*$ and $g_{*s}$ are the effective degrees of freedom for the energy and entropy densities, respectively.
When the temperature is high enough, such as larger than $O(100)$~GeV, these two effective degrees of freedom get close to each other around $g_*\sim g_{*s} \sim 100$.
It should be noted that $w$ is independent of the temperature during the radiation dominated period.
The Eq.~(\ref{Boltzmann1}) can be further simplified by introducing $\m_L/T\equiv \xi_Le^{wT}$,
which can be simply solved by
\begin{eqnarray}
	\xi_L(T) = \xi_L(T_{ini})-\a w \int^{T}_{T_{ini}} dT'\, \frac{\m_\chi(T')}{T'}e^{-wT'}, 
	\label{Boltzmann2}
\end{eqnarray}
where $T_{ini}$ denotes the initial temperature to solve the Boltzmann equation.
Here, $T_{ini}$ takes to be larger than $T_{osc}$ whose expression is shown in Eq.~(\ref{eq:Tosc and TBL}).
Since $\m_\chi=0$ (and thus $\xi_L=0$) at $T=T_{ini}$, our results shown in the next section do not depend on the value of $T_{ini}$.
After the decoupling of the lepton violating process, i.e. $w T \ll 1$, the lepton asymmetry  
ends up with $\m_L/T = \xi_L(T)$.

\subsection{Numerical results}
\label{sec:result}
Our goal in this section is to search for viable parameter regions for successful Leptogenesis.
After inflation, the Majoron is settled at its initial position, and hence, 
the dynamical level splittings are vanishing in that period, i.e. $\lim_{t\to0}\m_\chi(t)\propto\lim_{t\to0}\dot\chi(t)=0$.
Besides, the chemical potentials after inflation are expected to be zero, since any
asymmetry before inflation has been diluted away by inflation, i.e. $\mu_L = \mu_Q = \mu_H = 0$.
With these initial conditions, we solve the Boltzmann equation Eq.\,(\ref{Boltzmann2}) 
by the time that the temperature of the universe decreases to 
the sphaleron decoupling temperature $T_{\rm sph}\sim 100$ GeV.
Around the temperature $T_{\rm sph}$, the sphaleron rate is sufficiently dumped, and the baryon abundance freezes, 
which is determined by the lepton abundance at that time through $n_B=-(52/93)n_L$.
Notably, free parameters of new physics in Eq.~(\ref{Boltzmann1}) are only $m_\chi$ and  $M_R$.
Thus, by remembering that the initial amplitude is of the order of $M_R$,
the baryon abundance is given as a function of $m_\chi$ for  given neutrino masses.

\begin{figure}[tbp]
	\begin{center}
		\includegraphics[width=.48\linewidth]{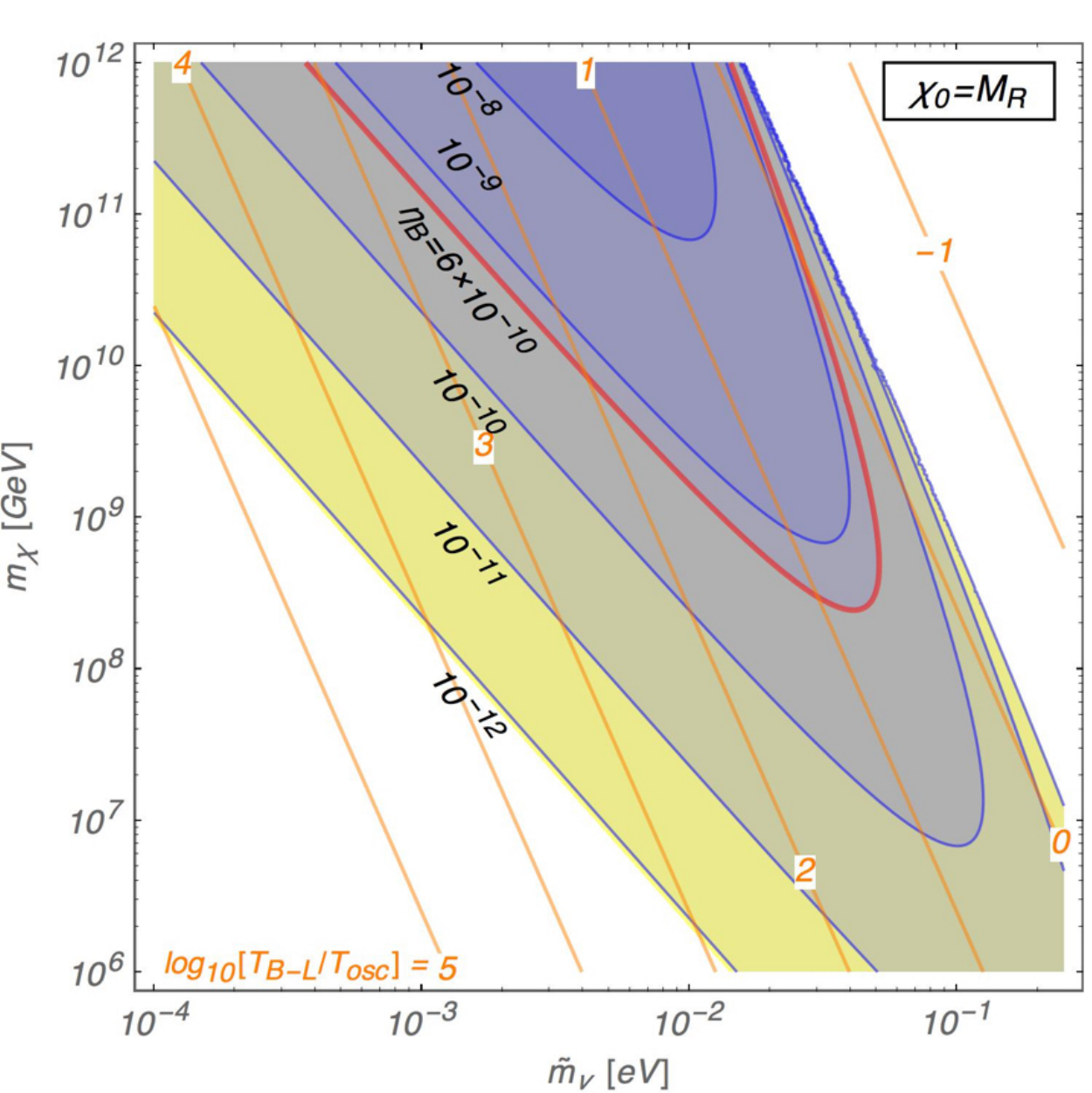}
		\hspace{10pt}
		\includegraphics[width=.48\linewidth]{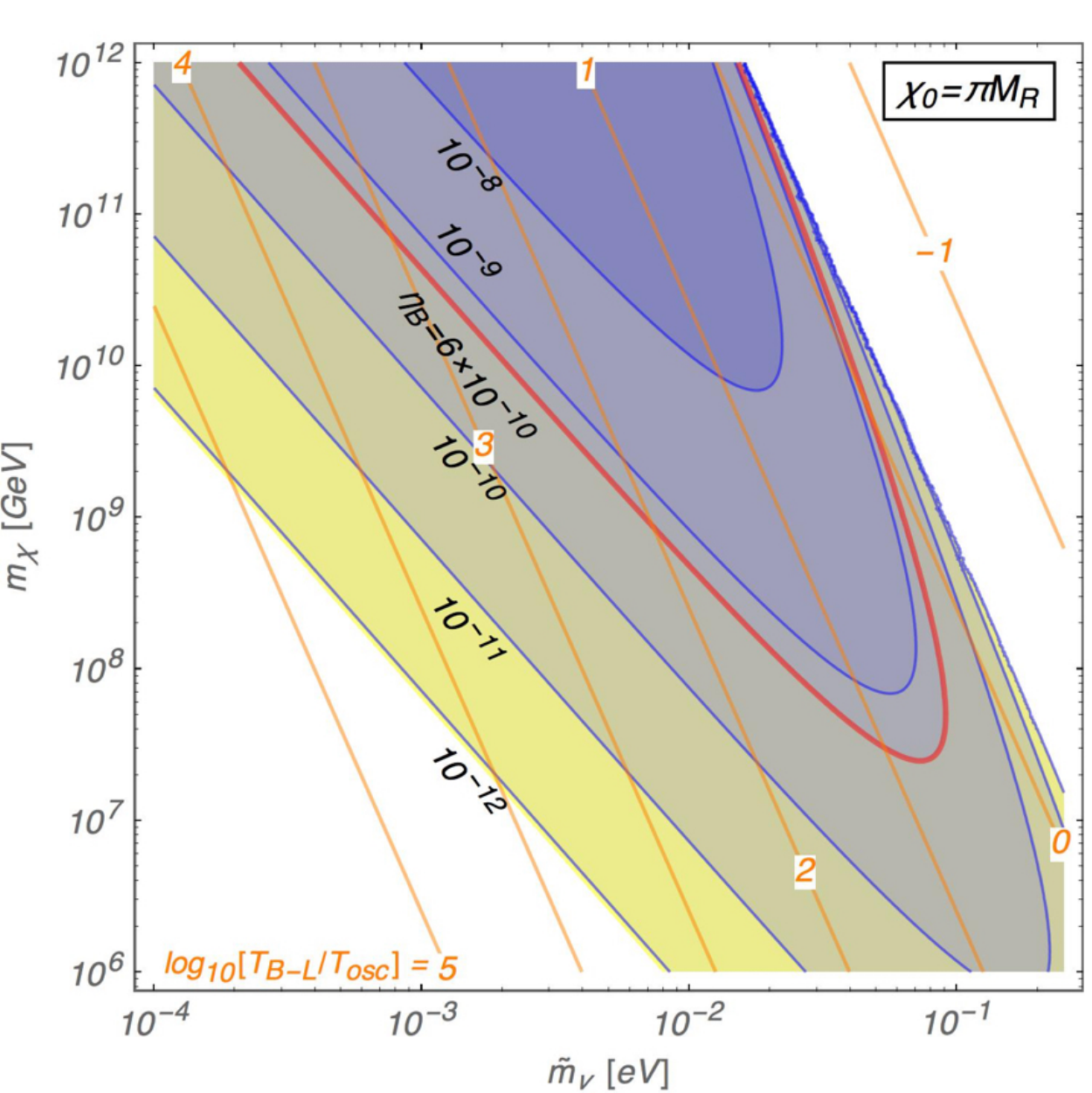}
		\caption{\sl\small The baryon-to-photon ratio $\eta_B=n_B/n_\g$ at today's temperature is shown as a function of $m_\chi$ and $\tilde m_\n$, where $\tilde m_\n^2\equiv \sum_i m_{\n_i}^2$. The orange lines represent $\log_{10}[T_{B-L}/T_{osc}]$ whose temperatures are defined in the text. The left panel of the figure corresponds to the case where the initial amplitude of Majoron field is $\chi_0=M_R$, whereas $\chi_0=\pi M_R$ is shown in the right panel.}
		\label{fig_scan}
	\end{center}
\end{figure}

Figure \ref{fig_scan} shows the contours of $\eta_B=n_B/n_\g$ at today's temperature as a function of 
$m_\chi$ and $\tilde m_\n$, where $\tilde m_\n^2$ is defined by the squared sum of neutrino masses, 
$\tilde m_\n^2\equiv \sum_i m_{\n_i}^2$. 
Here, we are taking the mass diagonal basis of the three neutrinos, so that $m_{\n_i}$'s  
 appear in the coefficients of the dimension five operator in Eq.~(\ref{Leff}).
In the figure, we have assumed $\chi_0=M_R$ (left panel) and $\pi M_R$ (right panel) as typical values.
Since the wash-out factor is proportional to $\tilde m_\n^2$, the net baryon asymmetry is strongly washed out 
when $\tilde m_\n$ is large.
On the other hand, if $\tilde m_\n$ is small, sufficient lepton asymmetry cannot be achieved since the lepton violating
processes are necessary to generate the asymmetry from the level splittings. 
These behaviors of the lepton (baryon) asymmetry can be also understood by comparing the temperatures, 
$T_{osc}$ and $T_{B-L}$, at which the Majoron starts to oscillate, $H(T_{osc})=m_\chi$, 
and the lepton number violating interactions decouple 
from thermal bath, $H(T_{B-L})=\langle\s_0v\rangle T_{B-L}^3$, respectively.
Suppose the universe is dominated by the radiation, we obtaion
\begin{eqnarray}
	T_{osc} \simeq 1.7\times g_*^{-1/4}(m_\chi M_{\rm Pl})^{1/2},
	&\quad&
	T_{B-L} \simeq 0.33\times g_*^{1/2}\langle\s_0 v\rangle^{-1}/M_{\rm Pl},
	\label{eq:Tosc and TBL}
\end{eqnarray}
where they are typically $T_{osc}\sim10^{13}$ GeV for $m_\chi\sim10^{9}$ GeV and $T_{B-L}\sim10^{13}$ GeV for $\langle\s_0v\rangle\sim10^{-31}~{\rm GeV}^{-2}$.
The orange lines in Figs.~\ref{fig_scan} show $\log_{10}[T_{B-L}/T_{osc}]$.
In the region where $\tilde m_\n$ is large,  the lepton number violating interactions strongly couple 
to the thermal bath, and the lepton asymmetry is significantly washed out.
It turns out that when $T_{B-L}$ is lower than $T_{osc}$, it is hard to generate enough lepton asymmetry.
In the region where $\tilde m_\n$ is small, on the other hand, $T_{B-L}$ is much higher than $T_{osc}$, and thus, the lepton number violating interaction is not effective when the Majoron field starts to oscillate.
As we will see, $\tilde m_\n$ is bounded below by experiments.
We therefore obtain upper limit on $\tilde m_\n$ for successful Leptogenesis.
As a result, we find the allowed neutrino mass ranges for $\eta_B\sim 6\times10^{-10}$, 
\begin{eqnarray}
	&\tilde m_\n& \lesssim 5.5\times 10^{-2}~{\rm eV},~~~~(\chi_0=M_R),\label{NH_range}\\
	&\tilde m_\n& \lesssim 9.1\times 10^{-2}~{\rm eV},~~~~(\chi_0=\pi M_R).
	\label{IH_range}
\end{eqnarray}
It should be noted that the results are saddled with $O(10)$\% error caused by the Maxwell-Boltzmann approximation.

\subsection{Neutrinoless double beta decay}
The allowed ranges of $\tilde m_\n$ can be translated into the allowed ranges of the neutrino spectrum by taking neutrino mass orderings into account, i.e. the normal hierarchy (NH) or the inverted hierarchy (IH)\footnote{Quasi-degenerate spectrum is also possible experimentally. However, such spectrum is not favored for successful Leptogenesis in our scenario since wash-out effect becomes too strong.}.
The NH spectrum corresponds to $m_{\n_3}>m_{\n_2}>m_{\n_1}$, whereas the IH spectrum to $m_{\n_2}>m_{\n_1}>m_{\n_3}$.
The observed values of mixing angles and squared mass differences are given by
\begin{eqnarray}
	&&
	\sin^2(2\theta_{12}) = 0.846,~~~\sin(2\theta_{13}) = 9.3\times 10^{-2},~~~\sin(2\theta_{23}) = 1.0,\nonumber\\
	&&
	m_S^2\equiv m_{\n_2}^2-m_{\n_1}^2 = 7.54\times10^{-5}~{\rm eV}^2,\nonumber\\
	&&
	m_A^2\equiv|m_{\n_3}^2-m_{\n_1}^2|=2.47\times10^{-3}~{\rm eV}^2~~(NH),~~~2.39\times10^{-3}~{\rm eV}^2~~(IH),
\end{eqnarray}
where we have taken the central values given in Ref.~\cite{Agashe:2014kda}.
Then, we have two mass spectra as
\begin{eqnarray}
	\text{NH}:&&
	m_{\n_1}\equiv m_0,~m_{\n_2}=\sqrt{m_0^2+m_S^2},~m_{\n_3}=\sqrt{m_0^2+m_A^2},\\
	\text{IH}:&&
	m_{\n_1}=\sqrt{m_0^2+m_A^2},~m_2=\sqrt{m_0^2+m_S^2+m_A^2},~m_{\n_3}\equiv m_0,
\end{eqnarray}
where $m_0$ denotes the lightest neutrino mass in each mass ordering.

\begin{figure}[tbp]
	\begin{center}
		\includegraphics[width=.49\linewidth]{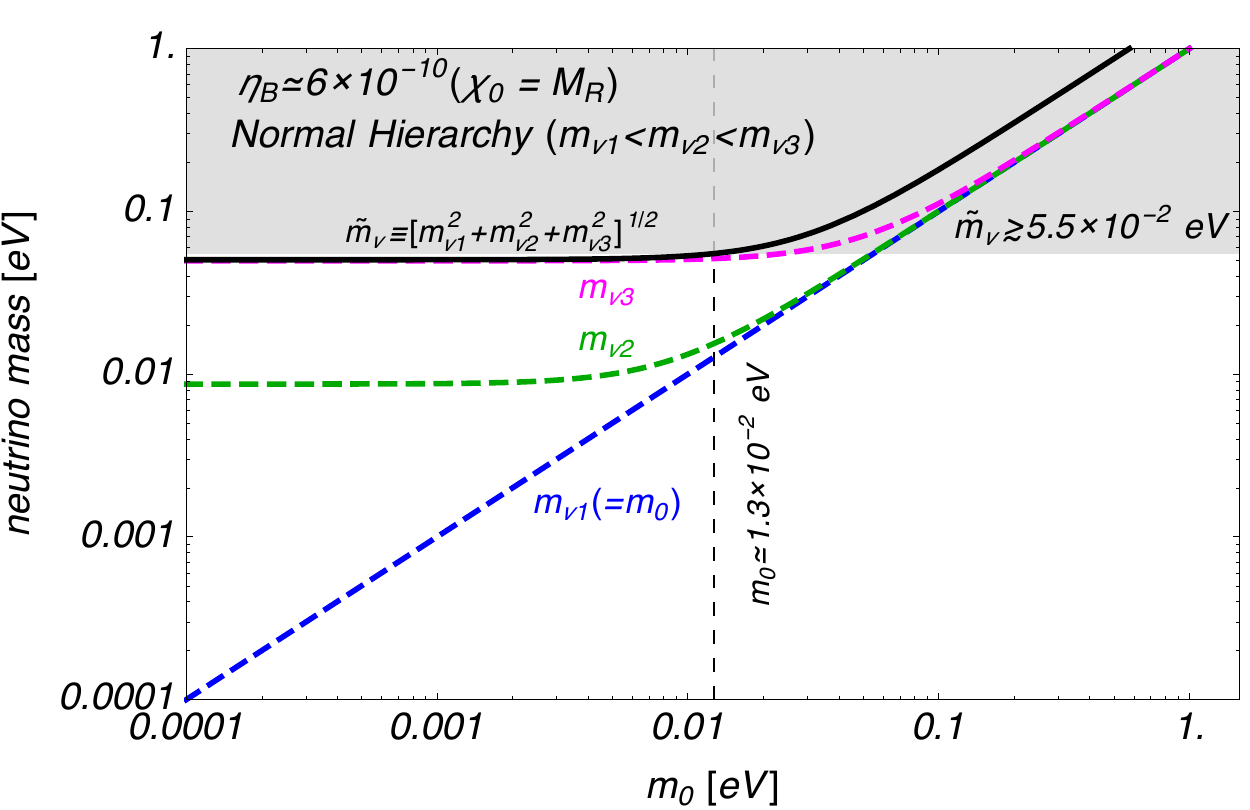}
		\hspace{1pt}
		\includegraphics[width=.49\linewidth]{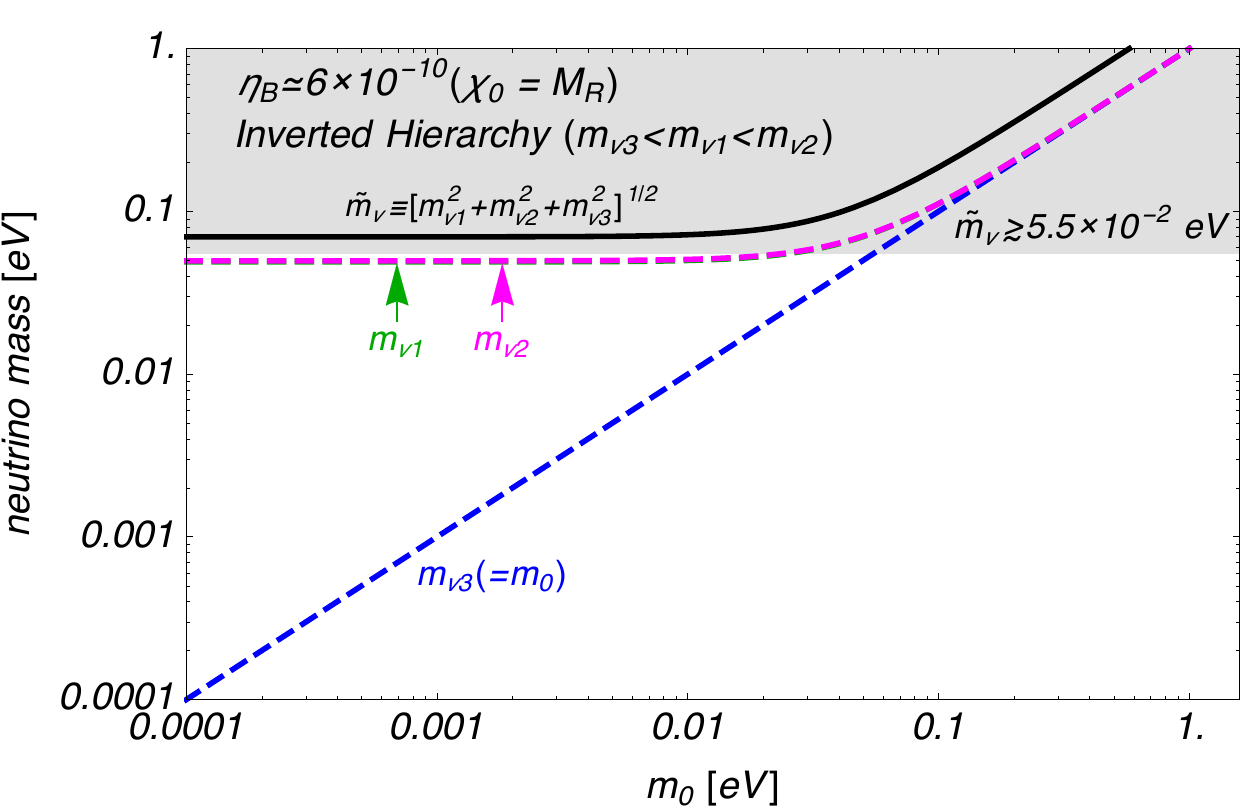}\\
		\includegraphics[width=.49\linewidth]{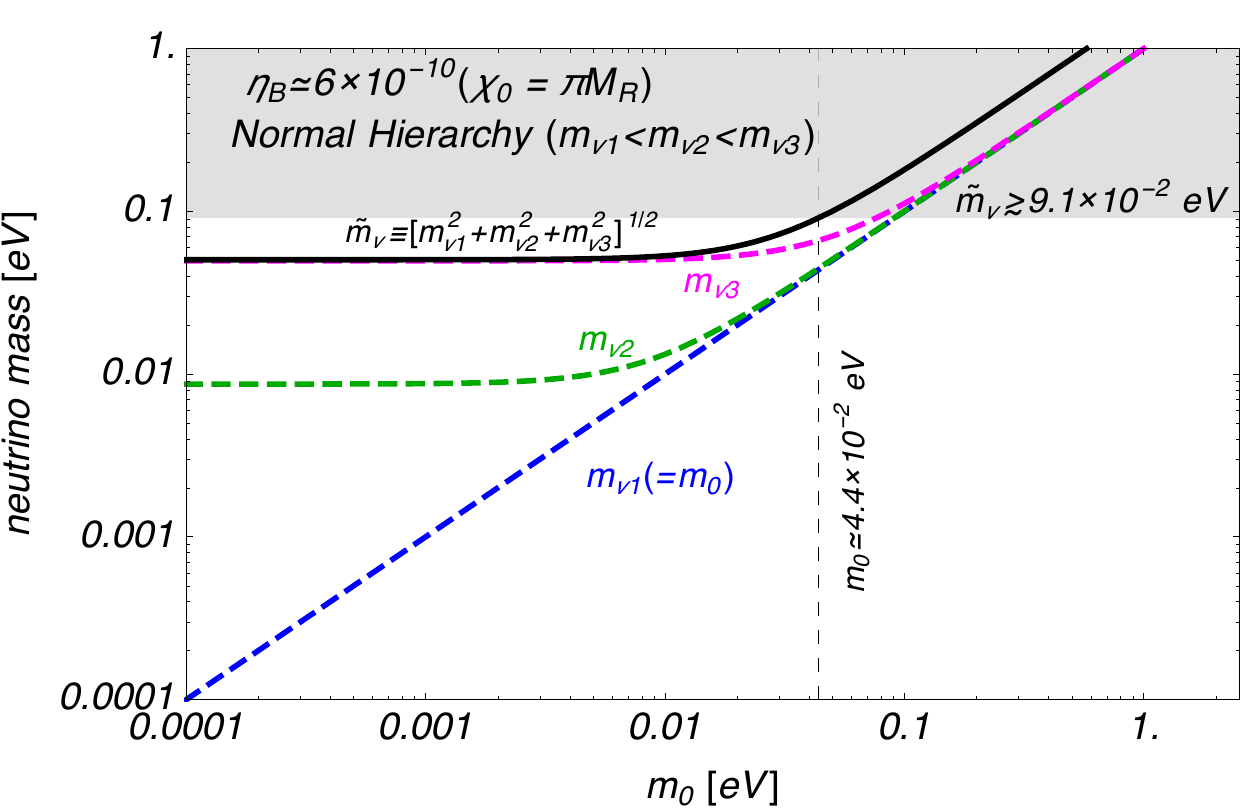}
		\hspace{1pt}
		\includegraphics[width=.49\linewidth]{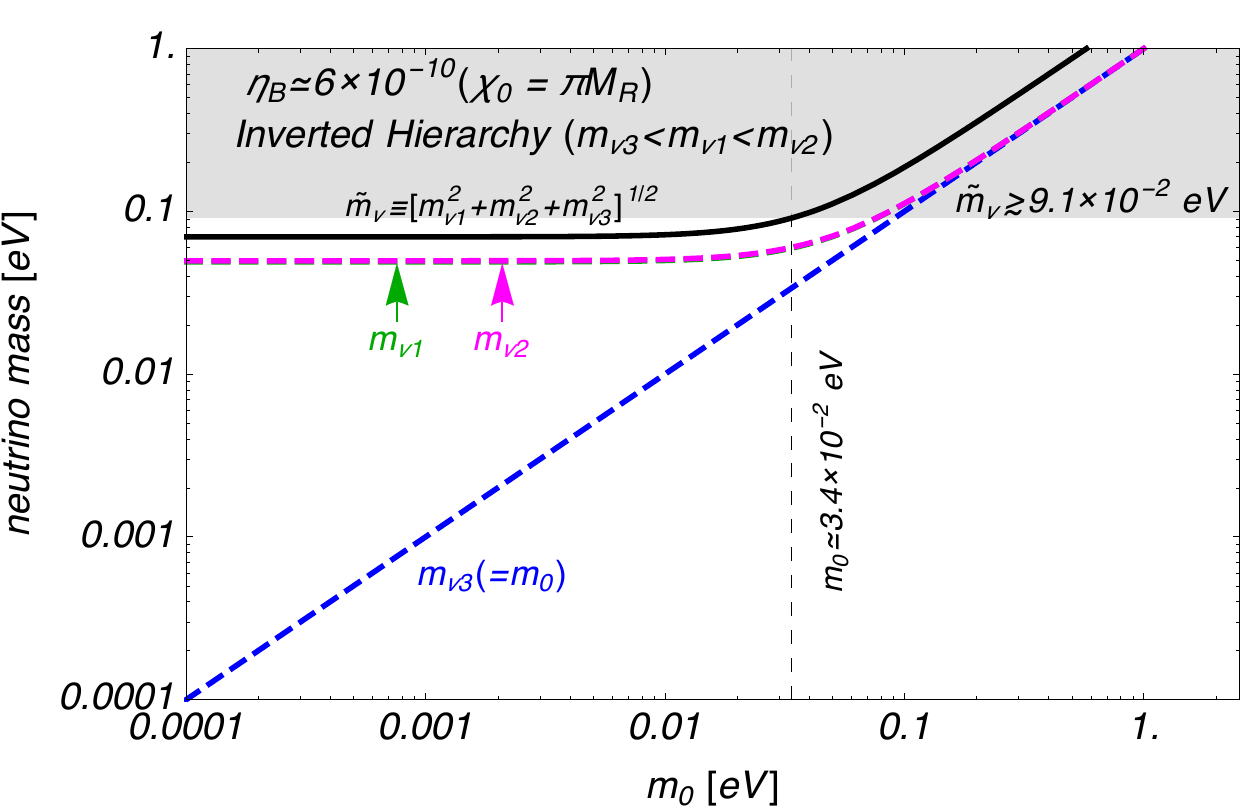}
		\caption{\sl\small Possible neutrino mass ranges are shown as a function of $m_0$, the lightest neutrino mass. The initial Majoron amplitude is taken to be $\chi_0=M_R$ in the upper two panels of the figure, whereas the bottom two panels show the case of $\chi_0=\pi M_R$. The left two panels are assumed the NH neutrino spectrum, whereas the IH spectrum is considered in the right two panels. The gray regions are disfavored since neutrino masses are too heavy/small to produce sufficient lepton asymmetry.}
		\label{fig_mnu}
	\end{center}
\end{figure}

Figure \ref{fig_mnu} shows favored regions of neutrino masses for successful Leptogenesis in our scenario, where $m_{\n_1},~m_{\n_2},~m_{\n_3}$ 
and $\tilde m_\n$ are depicted by three dashed lines and black solid line as a function of $m_0$.
In the upper two panels of Fig.~\ref{fig_mnu}, the initial Majoron amplitude is taken to be $\chi_0=M_R$, 
whereas the bottom two panels represent the cases for $\chi_0=\pi M_R$.
The left two panels are assumed the NH neutrino spectrum, whereas the IH spectrum is considered in the right two panels.
In all the panels of Fig.~\ref{fig_mnu}, the gray shaded regions depict the disfavored regions for successful Leptogenesis, 
which correspond to the ranges in Eqs.~(\ref{NH_range}) and (\ref{IH_range}).
Therefore, the allowed regions exist only in the case that the line of $\tilde m_\n$ comes into the white region between two gray shed area, which is almost determined by the atmospheric neutrino mass scale $m_A$.
In the case of $\chi_0=M_R$ the IH spectrum is rather disfavored in all parameter regions, 
while sufficient lepton asymmetry can be achieved in the NH spectrum with the parameter region of $m_0\lesssim1.3\times10^{-2}$ eV.%
\footnote{Here, again it should be cautioned that  we have $O(10)$\% errors in our estimations.}
On the other hand, in the case of $\chi_0=\pi M_R$, both the NH and the IH  spectra are consistent with
successful Leptogenesis where $m_0\lesssim4.4\times10^{-2}~{\rm eV}$ for the NH spectrum and $m_0\lesssim3.4\times10^{-2}~{\rm eV}$ for the IH spectrum.

\begin{figure}[tbp]
	\begin{center}
		\includegraphics[width=.5\linewidth]{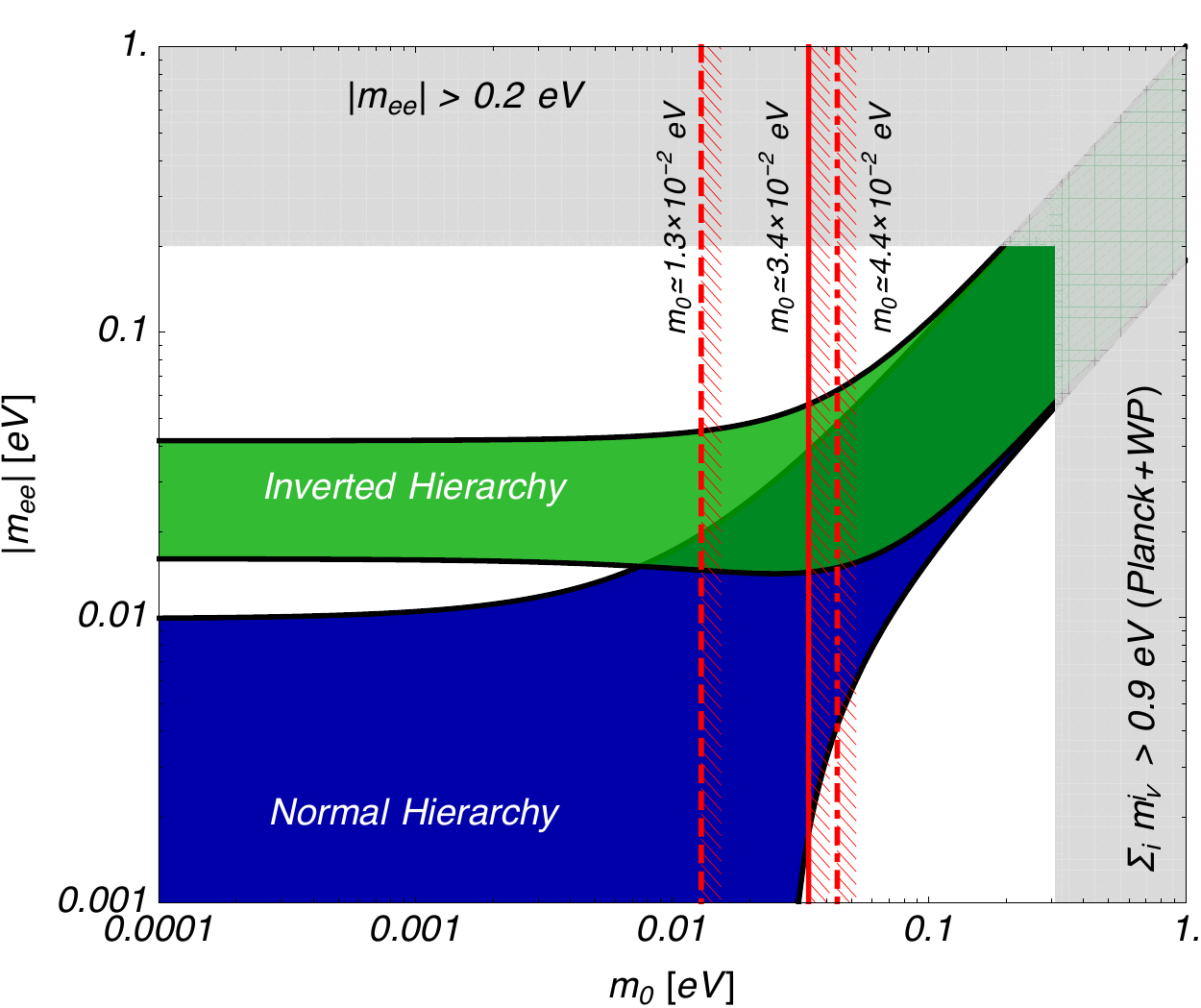}
		\caption{\sl\small The effective mass, $|m_{ee}|$, is shown as a function of $m_0$, where the NH and the 
		IH cases are depicted by the blue and the green regions, respectively. The upper bound for $m_0$ comes from the CMB observations, and the upper bound for $|m_{ee}|$ is given by Refs.~\cite{Albert:2014awa,Asakura:2014lma}. The red dashed line shows the limit for the NH spectrum in the case of $\chi_0=M_R$. The red dot-dashed and the solid lines put the upper limits on the NH and the IH cases for $\chi_0=\pi M_R$, respectively.}
		\label{fig_nu0bb}
	\end{center}
\end{figure}

The obtained constraints have  implications on the neutrinoless double beta decay whose decay width 
is proportional to a so-called effective mass, $|m_{ee}|^2$, 
\begin{eqnarray}
	|m_{ee}| &\equiv& \left|\sum_i m_{\n_i}U_{ei}^2\right|.
\end{eqnarray}
Here, $U$ denotes the Maki-Nakagawa-Sakata matrix parametrized by
\begin{eqnarray}
	U &=&
	\left(
	\begin{array}{ccc}
		1 & 0 & 0 \\
		0 & c_{23} & s_{23} \\
		0 & -s_{23} & c_{23}
	\end{array}
	\right)
	\left(
	\begin{array}{ccc}
		c_{13} & 0 & s_{13}e^{-i\delta} \\
		0 & 1 & 0 \\
		-s_{13}e^{i\delta} & 0 & c_{13}
	\end{array}
	\right)
	\left(
	\begin{array}{ccc}
		c_{12} & s_{12} & 0 \\
		-s_{12} & c_{12} & 0 \\
		0 & 0 & 1 
	\end{array}
	\right)
	\nonumber\\
	&& \times
	\left(
	\begin{array}{ccc}
		e^{i\alpha_1/2} & 0 & 0 \\
		0 & e^{i\alpha_2/2} & 0 \\
		0 & 0 & 1 
	\end{array}
	\right)\ ,
\end{eqnarray}
where $s_{ij}\equiv\sin(\theta_{ij}),~c_{ij}\equiv\cos(\theta_{ij})$, and $\delta$ and $\alpha_i$ are Dirac and Majorana phases, respectively.
Once we input a neutrino mass spectra, we obtain the effective mass as a function of $m_0$ as shown in Fig.~\ref{fig_nu0bb}, where the Majorana phases are scanned for $[0,2\pi]$.
It should be noted that though the effective mass also depends on the Dirac phase, the dependence 
is degenerated with those of the Majorana phases.
The upper limit on $m_0$ comes from the CMB observation, $\sum_i m_{\n_i}\lesssim 0.9$\,eV~\cite{Planck:2015xua}, 
which is roughly the same for both the NH and the IH cases.
The range of effective mass $|m_{ee}|>0.2$~eV has been excluded 
by the null observations of the neutrinoless double beta decay \cite{Albert:2014awa,Asakura:2014lma}.

In Fig.~\ref{fig_nu0bb}, we show the upper limits on $m_0$ for the successful Leptogenesis obtained in Fig.~\ref{fig_mnu}.
In the case of $\chi_0=M_R$, the IH spectrum is not successful, and hence, only the NH case is shown, i.e. $m_0\lesssim1.3\times10^{-2}$\,eV as a red dashed line.
In the case of $\chi_0=\pi M_R$,  the limits, $m_0\gtrsim4.4\times10^{-2}$ eV (NH) and $m_0\gtrsim3.4\times10^{-2}$ eV (IH) are depicted by the red dot-dashed and solid lines, respectively.
Future measurements of neutrinoless double beta decay are expected to reach $|m_{ee}|=O(10)$~meV \cite{Gomez-Cadenas:2015twa}.
The figure shows that if $|m_{ee}|>O(10)$~meV is confirmed by the neutrinoless double beta decay,
almost all the parameter region for the successful Leptogenesis will be excluded.

\subsection{Viable models}

Finally let us discuss viable models consistent with cosmological observations.
Since the Majoron we are interested in is super heavy, $m_{\chi} \gtrsim 10^{10}$\,GeV,
Planck suppressed operators play important roles not only on their masses as in Eq.\,(\ref{Omass})
but also on its decay rate.
Here, we consider the following operators as examples:
\begin{eqnarray}
	{\cal O}_D^{(n)} &=& \frac{\s^{n-2}|H|^2}{M_{\rm Pl}^{n-4}}~~~(n=5,6,7\cdots) ,
	\label{Odecay}
\end{eqnarray}
to induce a rather large decay rate of the Majoron.%
\footnote{The decay rate via the derivative couplings is proportional to the neutrino masses,
and hence, very small.}
It should be noted that the Majoron masses generated by these operators are negligibly small compared to $O_M^{(n)}$ 
in Eq.~(\ref{Omass}). 
As discussed below, the power of $\s$ in ${\cal O}^{(n)}_D$ is expected to be the same as that in ${\cal O}^{(n)}_M$.
Therefore, let us suppose that the Majoron decay is induced by ${\cal O}^{(7)}_D$ for example, where Eq.~(\ref{Omass}) gives ${\cal O}^{(5)}_M$ to the Majoron mass in this case.
On the other hand, if ${\cal O}^{(7)}_D$ dominates the Majoron mass term at $H\simeq m_\chi$, Majoron would dissipate before the time to oscillate.
Let us derive the condition to evade such undesired case:
\begin{eqnarray}
	m_\chi^2\chi^2 &>& 
	\chi^5|h|^2/M_{\rm Pl}^3
	\sim (\chi^5/M_{\rm Pl})(T^2/M_{\rm Pl})
	\sim (\chi^5/M_{\rm Pl}) H_{osc}.
\end{eqnarray}
At the time to oscillate, we obtain
\begin{eqnarray}
	\chi_0 &<& (M_{\rm Pl}^2m_\chi)^{1/3}
	\sim 2.2\times 10^{15}
	\left[
	\left(
	\frac{M_{\rm Pl}}{10^{18}~{\rm GeV}}
	\right)^2
	\left(
	\frac{m_\chi}{10^{10}~{\rm GeV}}
	\right)
	\right]^{1/3} ~{\rm GeV}
\end{eqnarray}
by using $H_{osc}=m_\chi$ and $\chi=\chi_0$.
We can therefore avoid such situation in most cases.

Since we aim to discuss a connection to a possible ultraviolet completion of our model, 
we constrain ourselves on the cases where both the operators ${\cal O}_M$ and ${\cal O}_D$ are induced 
by a condensate of either a scaler field or a composite field.
For example, suppose a scalar filed $\Phi$ with charge $q_\Phi$ under the gauged $U(1)_{B-L}$ acquires a VEV, 
$\langle \Phi \rangle=v_\Phi$, corresponding ${\cal O}_M^{(n)}$ and ${\cal O}_D^{(n)}$ are induced from the operators such as
\begin{eqnarray}
	\tilde {\cal O}_M = \frac{\Phi}{M_*}\frac{\s^{q_\Phi/2}}{M_{\rm Pl}^{q_\Phi/2-4}},~
	\tilde {\cal O}_D = \frac{\Phi}{M_*}\frac{\s^{q_\Phi/2}|H|^2}{M_{\rm Pl}^{q_\Phi/2-2}},
\end{eqnarray}
where we assume $q_\Phi\geq10$, and $M_*$ is a certain high energy scale below $M_{\rm Pl}$.
The charge $q_\Phi$ is a model dependent parameter, and we demonstrate the cases of $q_\Phi=10$ and $12$ in this section,
leading to $Z_{10}$ or $Z_{12}$ $B-L$ symmetries, respectively. 
Eventually, spontaneous breaking of the resultant discrete symmetry induces Majoron, as we have discussed in the previous section.
It should be noted that in these cases the initial amplitude of Majoron is restricted to smaller range than $(0,\pi]\times v_{B-L}$.
For example, in the $Z_{10}$ model the remnant discrete symmetry is $Z_2$ after $\s$ acquires a VEV, and hence the Majoron takes an initial field value within the range of $\chi/\sqrt{2}=(0,\pi/5]\times v_{B-L}$.
The resultant baryon number asymmetry is, therefore, roughly the same as the left panel of the Fig.~\ref{fig_scan} when $\chi_0$ takes the maximal value, $\chi_0=\pi/5v_{B-L}\sim M_R$.
In this case only NH spectrum is favored as shown in Fig.~\ref{fig_mnu}.
Such restriction is rather relaxed in the $Z_{12}$ model.

\begin{figure}[tbp]
	\begin{center}
		\includegraphics[width=.48\linewidth]{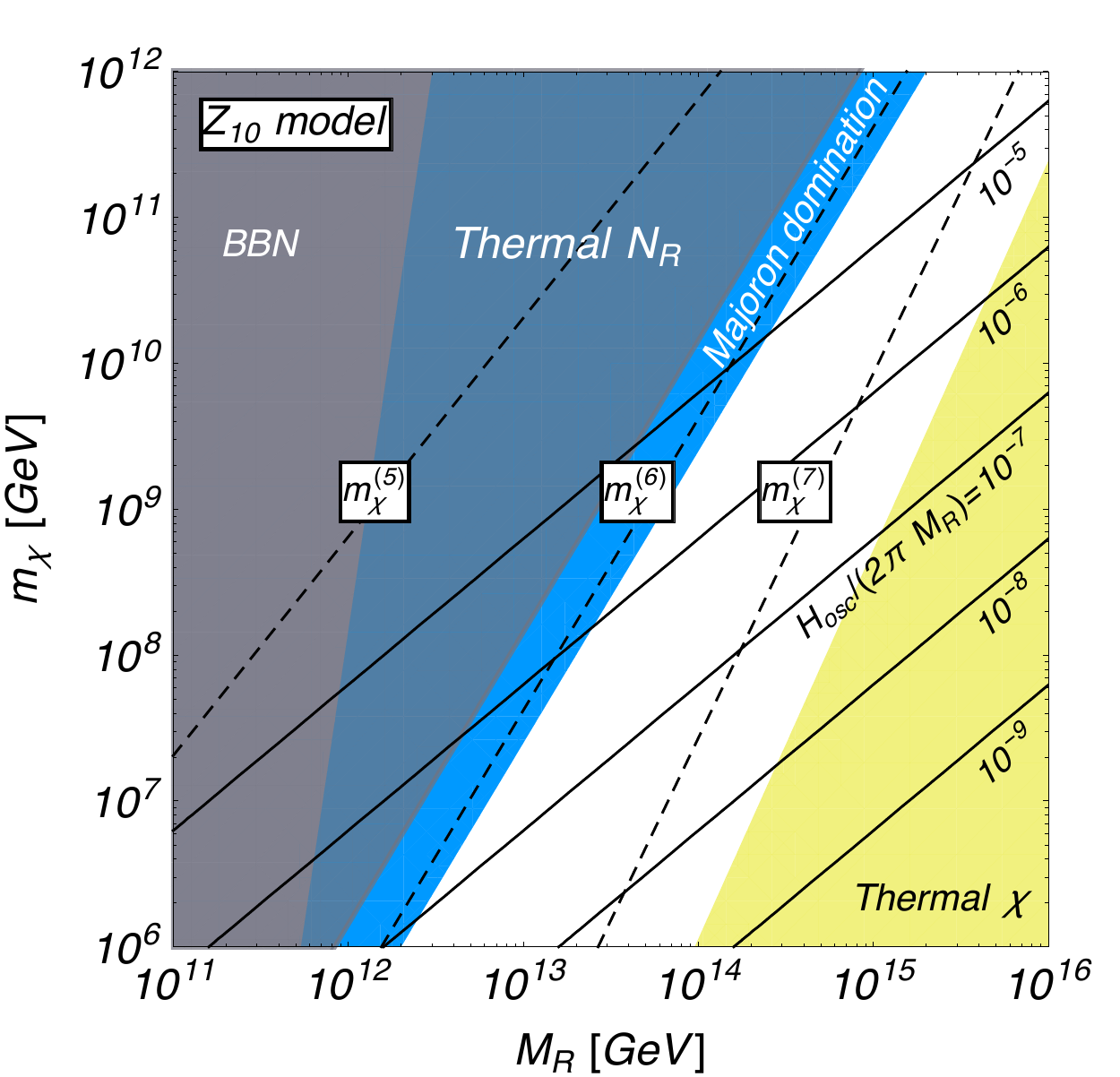}
		\hspace{10pt}
		\includegraphics[width=.48\linewidth]{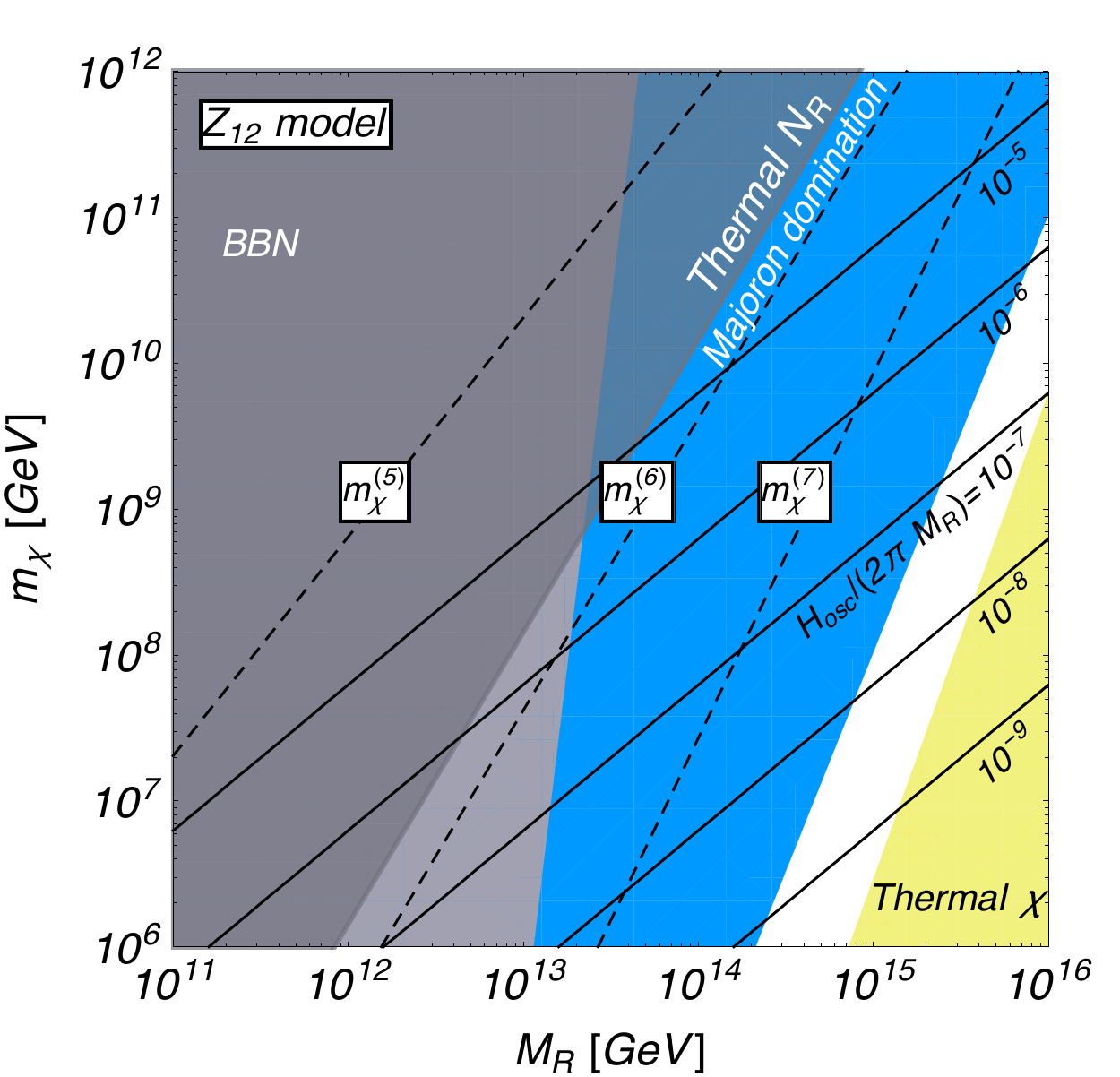}
		\caption{\sl\small Several cosmological bounds are shown in the $m_\chi$-$M_R$ plane. The left and right panels show the $Z_{10}$ and $Z_{12}$ models, respectively. The Majoron never enters the thermal bath other than the yellow shaded regions. In the blue regions the energy of the Majoron oscillation dominates the energy density of the universe. The deeper blue regions represent the range where right-handed neutrino is thermalized after reheating. In the gray region Majoron survive until BBN takes place. The black solid lines show the baryonic isocurvature fluctuations, and the black dashed lines depict the Majoron masses, $m_\chi^{(n)}$, given by Eq.~(\ref{mchi}).}
		\label{fig_model}
	\end{center}
\end{figure}

Figure~\ref{fig_model} shows several constraints on $m_\chi$ and $M_R$ for two example cases of $Z_{10}$ (left panel) and $Z_{12}$ (right panel), respectively.
In both cases, we require that the discrete symmetry is broken before inflation to avoid the domain wall problem.
In adittion, we also require that the Majoron starts to oscillate well after inflation.
Thus, the Hubble parameter during inflation, $H_{\rm inf}$, is much larger than the Majoron mass. 
In such situation, the quantum fluctuation of Majoron field causes a baryonic isocurvature perturbation, 
which is constrained by CMB observations.
In order to be consistent with the observations, the Majoron fluctuation should satisfy \cite{Turner:1988sq,Sasaki:1991ik,Harigaya:2014tla}: $10^{-5}\gtrsim\d\chi/\chi_0\sim H_{\rm inf}/(2\pi M_R)\gtrsim H_{osc}/(2\pi M_R)$,
 where $H_{osc}\equiv H(T_{osc}) \simeq m_{\chi}$. 
From this argument, at least, the condition $H_{osc}/(2\pi M_R)\lesssim 10^{-5}$ should be satisfied.
In the figure, we show the value of $H_{osc}/(2\pi M_R)$
as the black solid lines.

In the case of $Z_{10}$, the Majoron mass comes from ${\cal O}_M^{(5)}$, i.e. $m_\chi^{(5)}$ in Eq.\,(\ref{mchi}), 
while the $\chi$-$H$-$H$ interaction is induced from ${\cal O}_D^{(7)}\sim(M_R^4/M_{\rm Pl}^3)\chi|H|^2$.
It is necessary to pay attention whether the Majoron dominates the primordial energy density of radiation or not
before its decay, since the Majoron decay might dilute the generated lepton asymmetry.
The temperature at which the Majoron dominates, $T_{dom}$, depends on whether the Majoron is thermalized or not:
\begin{eqnarray}
	T_{dom} &=&
	\frac{15\z(3)}{\pi^4}\frac{m_\chi}{g_*}~~~\text{(thermal)},~~~\frac{M_R^2}{3M_{\rm Pl}^2}T_{osc}~~~\text{(non-thermal)}.
	\label{Tdom}
\end{eqnarray}
The Majoron thermalization occurs via the inverse decay process induced by
the $\chi$-$H$-$H$ coupling,%
\footnote{It should be noted that the Majoron couplings to the Standard Model particles
via the dimension five operators are highly suppressed by small neutrino masses.}
where the inverse decay rate is given by $\G_{th}\sim(M_R^4/M_{\rm Pl}^3)^2/(8\pi)/T$, 
which becomes equal to the Majoron decay $\G_D$ at $T \simeq m_{\chi}$.
The Majoron is therefore thermalized when $m_\chi<T_{th}$, 
where the thermalization temperature $T_{th}$ is obtained by solving the condition $H(T_{th})=\G_{th}(T_{th})$.

The yellow region in the left panel of Fig.~\ref{fig_model} satisfies $m_\chi<T_{th}$, where the Majoron 
is thermalized by the effect of the decay operator.
In that region, however, the domination temperature is always smaller than the decay temperature $T_D$ defined by
\begin{eqnarray}
	T_{D} &=& 
	\left(
	\frac{90M_{\rm Pl}^2\G_D^2}{\pi^4g_*}
	\right)^{1/4},
\end{eqnarray}
and thus, the dilution does not occur.
In the whole regions other than the yellow region, the Majoron does not enter the thermal bath, and the energy density of the Majoron oscillation dominates at $T_{dom}$ as given by Eq.~(\ref{Tdom}).
When $T_{dom}$ becomes larger than $T_D$, the Majoron dominates the universe, and causes dilution of baryon asymmetry 
due to entropy production, which is represented by the blue shaded region in the figure.
The gray shaded region in the figure shows that $T_D$ is smaller than the temperature at which BBN takes place, $T\sim O(1)$ MeV, where the Majoron decay spoils the successful BBN.

When $T_{osc}> M_R$, the right-handed neutrinos are in the thermal bath 
when spontaneous Leptogenesis takes place, which is depicted by the deeper blue shaded region in the figure.
In this region, conventional thermal Leptogenesis can take place if it is possible, and further lepton asymmetry could be generated even if the dilution occurs.
Since this possibility is out of our study, however, we do not discuss further.

Let us finally mention the Majoron mass $m_\chi^{(n)}$ represented by the black dashed lines in the figure.
In the $Z_{10}$ model, for example, the Majoron mass is obtained by $m_\chi^{(5)}$, and the line representing $m_\chi^{(5)}$ in the left panel of the figure is in the region where Majoron dominates the energy density of the universe.
This shows that we need $O(1$--$0.1)$\% fine-tuning of the coefficient of  ${\cal O}_M^{(5)}$
to obtain viable Majoron parameters for a given $M_R$.
The same figure for $Z_{12}$ shows that we need some amount of fine-tuning of the coefficient of  ${\cal O}_M^{(6)}$.

\section{Summary}
We have proposed a new type of spontaneous Leptogenesis caused by the oscillation of Majoron field.
As exploited by Cohen and Kaplan~\cite{Cohen1987,Cohen1988}, the mechanism evades the Sakharov's conditions.
Therefore, our scenario can work in the case that the conventional Leptogenesis does not work, e.g., even when there are not large enough $CP$ phases.
As a notable feature, all the necessary ingredients for spontaneous Leptogenesis are automatically equipped with in our setup, i.e., derivative couplings to the Majoron and the lepton number violating interactions.
Once the neutrino masses are determined, the resultant lepton asymmetry depends on the Majoron mass and the initial value of the Majoron amplitude.
On the other hand, the dynamical level splitting is induced by $\m_\chi$ which is normalized by $M_R$.
Since the initial amplitude of the Majoron field is typically to be $O(M_R)$, $\m_\chi$ is not strongly affected by $B-L$ symmetry breaking scale.
Therefore, the sufficient baryon asymmetry can be achieved if the Majoron has an appropriate mass.

To explain the observed baryon number asymmetry, we find that the neutrino masses are predicted in some ranges since the resultant lepton number asymmetry strongly depends on the wash-out effect caused by the 
dimension five operators of the neutrino masses.
As a result, we find that spontaneous Leptogenesis rather disfavors the degenerate neutrino spectrum.
The effective mass of neutrinoless double beta decay is also constrained accordingly.
We have also discussed viable models, which are $Z_{10}$ and $Z_{12}$ models, consistent with cosmological observations.

%%%%%%%%%%%%%%%%%%%%%%%%%%%%%%%%%%%%%%%
%%%%%%%%%%% Acknowledgments %%%%%%%%%%%
%%%%%%%%%%%%%%%%%%%%%%%%%%%%%%%%%%%%%%%
\section*{Acknowledgments}
We thank A.~Kusenko for useful discussion on the spontaneous Baryogenesis.
K.K. also thank M.~Yamada for useful discussions.
This work is supported by Grant-in-Aid for Scientific research from the Ministry of Education, Science, Sports, and Culture (MEXT), Japan, No. 24740151 and 25105011 (M.I.), from the Japan Society for the Promotion of Science (JSPS), No. 26287039 (M.I.), the World Premier International Research Center Initiative (WPI Initiative), MEXT, Japan (M.I.).

%%%%%%%%%%%%%%%%%%%%%%%%%%%%%%%%%%
%%%%%%%%%%% References %%%%%%%%%%%
%%%%%%%%%%%%%%%%%%%%%%%%%%%%%%%%%%

\appendix
\section{Cross sections in the presence of the Majoron background}
\label{sec:appendix1}

In this appendix we give detailed calculations of scattering cross sections in Majoron background.

\subsection{Fermions with non-vanishing $\m_\chi$}

Let us first consider a free Dirac fermion, which corresponds to the limit of the vanishing
dynamical level splittings, $\m_\chi\to0$ in Eq.~(\ref{Lkin}).
In this case, the Dirac field satisfies $(i\slashed{\del}-m)\p=0$, and then we expand the solution as 
\begin{eqnarray}
	& &
	\p(x) = \p_+(x) + \p_-(x),\\
	& &
	\p_+(x) = \int\frac{d^3p}{(2\pi)^3}b_{\vec p} u_{\vec p}(t) e^{i\vec p\cdot \vec x},~~~
	\p_-(x) = \int\frac{d^3p}{(2\pi)^3}d_{\vec p}^\dagger v_{\vec p}(t) e^{-i\vec p\cdot \vec x},
\end{eqnarray}
where we omit spin indexes.
The independent solutions $u_{\vec p}$ and $v_{\vec p}$ satisfy the Dirac equation for a particle and anti-particle, respectively, and hence, $b_{\vec p}$ and $d_{\vec p}$ are the creation and the annihilation operators of the particles and the anti-particles.
The time dependence of these solutions can be expressed by the energy eigenvalues, $\w_0(\vec p)\equiv\sqrt{|\vec p|^2+m^2}$, as follows:
\begin{eqnarray}
	u_{\vec p} = [2\w_0(\vec p)]^{-1}u_{\vec p}^0e^{-i\w_0(\vec p) x^0},~~~
	v_{\vec p} = [2\w_0(\vec p)]^{-1}v_{\vec p}^0e^{i\w_0(\vec p) x^0},
\end{eqnarray}
where we choose a normalization by recovering spin indexes $r$ and $s$ as
\begin{eqnarray}
	u^{0\dagger}_{\vec p, r}u^0_{\vec p, s}=2\w_0(\vec p)\d_{rs},~~~
	v^{0\dagger}_{\vec p, r}v^0_{\vec p, s}=2\w_0(\vec p)\d_{rs}.
\end{eqnarray}
As a result, we can expand $\psi$ in terms of $b_{\vec p}$ and $d_{\vec p}$,
\begin{eqnarray}
	\p(x) = \int\frac{d^3p}{(2\pi)^32\w_0(\vec p)}\left[b_{\vec p}u^0_{\vec p}e^{-ip\cdot x} + d^\dagger_{\vec p}v^0_{\vec p}e^{ip\cdot x}\right],
\end{eqnarray}
where we define $p=(\w_0,\vec p)$.
This leads to the momentum conservation of ${\cal S}$ matrix for 2-body process, for example, $h^+(p_1)h^+(p_2)\to e^+_L(p_3)e^+_L(p_4)$, as
\begin{eqnarray}
	{\cal S} = (2\pi)^4\d(p_1+p_2-p_3-p_4)\cdot i{\cal M},
	\label{S-matrix1}
\end{eqnarray}
where ${\cal M}$ is a scattering amplitude.

Next let us see the case of non-vanishing $\m_\chi$ where Eq.~(\ref{S-matrix1}) is slightly deformed.
In this case, the Lagrangian of a fermion is given by,
\begin{eqnarray}
	{\cal L} &=& \bar\p (i\slashed{\del}-m-\m_\chi\g^0)\p.
	\label{L_psi_chi}
\end{eqnarray}
As discussed in section \ref{sec:SSB and Majoron oscillation}, $\m_\chi$ stems from 
the Majoron background in motion, and thus it depends on time.
The situation of our interest is the case at which the Majoron starts to oscillate in the expanding universe.
The time scale of the Majoron oscillation is, therefore, $O(1/m_\chi)$.
The typical time scale of the scattering among $\p$'s is, on the other hand, characterized by $1/T$ since $\p$ is in the thermal bath.
The temperature at which the Majoron starts to oscillate is roughly obtained by $H=m_\chi$, and hence,
$1/T\sim 1/\sqrt{M_{\rm Pl} m_\chi}$ is much shorter than the time scale of Majoron oscillation.
We can therefore treat $\m_\chi$ as a constant in the cross section calculations.

In the similar way to the previous case, we can take the solution of the Dirac equation given by Eq.~(\ref{L_psi_chi}) as
\begin{eqnarray}
	& &
	\tilde\p(x) = \tilde\p_+(x) + \tilde\p_-(x),\\
	& &
	\tilde\p_+(x) = \int\frac{d^3p}{(2\pi)^3}b_{\vec p} \tilde u_{\vec p}(t) e^{i\vec p\cdot \vec x},~~~
	\tilde\p_-(x) = \int\frac{d^3p}{(2\pi)^3}d_{\vec p}^\dagger \tilde v_{\vec p}(t) e^{-i\vec p\cdot \vec x}.
\end{eqnarray}
It should be noted that these solutions, $\tilde u_{\vec p}$ and $\tilde v_{\vec p}$, have deformed dispersion relations given by $\w(\vec p)=\pm\sqrt{|\vec p|^2+m^2}+\m_\chi=\pm\w_0(\vec p)+\m_\chi$.
The time dependence of the solutions is therefore governed by $\w(\vec p)$, and hence we obtain
\begin{eqnarray}
	u_{\vec p} = [2\w_0(\vec p)]^{-1}u_{\vec p}^0e^{-i\m_\chi x^0}e^{-i\w_0(\vec p) x^0},~~~
	v_{\vec p} = [2\w_0(\vec p)]^{-1}v_{\vec p}^0e^{-i\m_\chi x^0}e^{i\w_0(\vec p) x^0}.
\end{eqnarray}
It should be noted that $u^0$ and $v^0$ are the one defined for $\m_\chi = 0$.
This result implies that   ${\cal M}$ does not depend on $\m_\chi$, and hence,
the ${\cal S}$ matrix depends on $\m_\chi$ only through the 
$\delta$-function of the four momentum, i.e.
\begin{eqnarray}
	{\cal S} = (2\pi)^4 \d^4(p_1+p_2-p_3-p_4+2p_\chi)\cdot i{\cal M},
	\label{modified S-matrix}
\end{eqnarray}
where $p_\chi\equiv (\m_\chi,\vec 0)$ in a certain frame.
%Therefore, the cross section depends on $\m_\chi$ only when 
%${\cal M}$ depends on the four momenta of the final state particles.

\subsection{$\m_\chi$ dependence of phase space integrations}

We now show that the phase space factor  does not depend on the effective chemical potential at the leading order of $\m_\chi$.
First, we again clarify our situation; we are focusing on that the universe is filled by a background field $\chi$ which is a function of time, and only $p_\chi=(\del_0\chi,\vec 0)$ takes place the effective chemical potential.
In this case the frame taking $\vec p_\chi=\vec 0$ is useful for our purpose.
However, to calculate the reaction rate, i.e., the phase space integration,
it is much easier to consider the center of mass frame of the initial state.
We therefore relate these two frames by Lorentz boosts.

Suppose the initial particles have momenta $p_1=(E_1,\vec p_1)$ and $p_2=(E_2,\vec p_2)$, and the initial total momentum is defined by $P_{\rm ini} = p_1+p_2$.
In general, when we are in the frame where $p_\chi=(\m_\chi,\vec 0)$, $P_{\rm ini}$ is not in its center of mass frame.
Here, we call this frame as the {\it original} frame where the momentum set is $\{P_{\rm ini}^{(O)}, p_\chi^{(O)}\}$.
Each momentum is written by
\begin{eqnarray}
	P_{\rm ini}^{(O)} &=& (\sqrt{S+|\vec P_{\rm ini}|^2},\vec P_{\rm ini}),\\
	p_\chi^{(O)} &=& (\m_\chi,\vec 0).
\end{eqnarray}
On the other hand, we can take the frame of $\vec P_{\rm ini}^{(O)}=\vec 0$, which we call the {\it center of mass} frame where the momentum set is given by $\{P_{\rm ini}^{(CMS)}, p_\chi^{(CMS)}\}$.
Each momentum is written by
\begin{eqnarray}
	P_{\rm ini}^{(CMS)} &=& (\sqrt{S},\vec 0),\\
	p_\chi^{(CMS)} &=& \frac{\m_\chi}{\sqrt{S}}\times(\sqrt{S+|\vec P_{\rm ini}|^2},-\vec P_{\rm ini}).
\end{eqnarray}
These two frames, $\{P_{\rm ini}^{(O)}, p_\chi^{(O)}\}$ and $\{P_{\rm ini}^{(CMS)}, p_\chi^{(CMS)}\}$, are transformed by the Lorentz boost each other: $P_{\rm ini}^{(CMS)}=\L^{-1}P_{\rm ini}^{(O)}$ and $p_\chi^{(CMS)}=\L^{-1}p_\chi^{(O)}$, where
\begin{eqnarray}
	\L &=&
	\left(
	\begin{array}{cc}
		\g & \g \vec\b \\
		\g \vec\b^T & \1+(\g-1)\frac{\vec \b^T\vec \b}{|\vec \b|^2}
	\end{array}
	\right),~~~
	\L^{-1} =
	\left(
	\begin{array}{cc}
		\g & -\g \vec\b \\
		-\g \vec\b^T & \1+(\g-1)\frac{\vec \b^T\vec \b}{|\vec \b|^2}
	\end{array}
	\right),\\
	\vec\b &=& \frac{\vec P_{\rm ini}}{\sqrt{S+|\vec P_{\rm ini}|^2}},~~~
	\g=\frac{1}{\sqrt{1-|\vec \b|^2}} = \frac{\sqrt{S+|\vec P_{\rm ini}|^2}}{\sqrt{S}}.
\end{eqnarray}
This allows us to obtain the cross section in the original frame after performing calculations in the center of mass frame.

In the center of mass frame we consider the following quantities:
\begin{eqnarray}
	d\F(p_3,p_4) &\equiv& \frac{d^3p_3}{(2\pi)^32E^0_3(\vec p_3)}\frac{d^3p_4}{(2\pi)^32E^0_4(\vec p_4)}(2\pi)^4\d^4(P_{\rm ini}-p_3-p_4+2p_\chi),\\
	I  &\equiv& \int d\F(p_3,p_4)\sum_{\rm spins}|{\cal M}(p_1,p_2,p_3,p_4)|^2,
\end{eqnarray}
where we take massless limit and $E^0_i(\vec p)\equiv\sqrt{|\vec p|^2}$ ($i=3,4$).
After reparametrizing $p_3-p_\chi\equiv \tilde p_3$ and $p_4-p_\chi\equiv \tilde p_4$, we immediately obtain $\vec{\tilde p}_4=-\vec{\tilde p}_3\equiv -\vec{\tilde p}$ due to $\vec P_{\rm ini}=\vec 0$.
The remaining delta function implies $E_{\rm tot}-E^0_3(\vec{\tilde p}_3-\vec p_\chi)-E^0_4(\vec{\tilde p}_4+\vec p_\chi)=0$ where $E_{\rm tot}\equiv \sqrt{S}+2p_\chi^{(CMS)0}$.
By eliminating this delta function, we obtain
\begin{eqnarray}
	|\vec{\tilde p}|^2 &=& \frac{E_{\rm tot}^2(E_{\rm tot}^2-4|\vec p_\chi^{(CMS)}|^2)}{4E_{\rm tot}^2-16|\vec p_\chi^{(CMS)}|^2(\vec{\tilde n}\cdot\vec n_\chi)},
\end{eqnarray}
where $\vec p_\chi^{(CMS)}\equiv|\vec p_\chi^{(CMS)}|\vec n_\chi$ and $\vec{\tilde p}\equiv |\vec{\tilde p}|\vec{\tilde n}$.
We finally obtain the expression for $d\F(\tilde p_3,\tilde p_4)$ as
\begin{eqnarray}
	d\F(\tilde p_3,\tilde p_4)
	&=&
	\frac{1}{4(2\pi)^2}\frac{|\vec{\tilde p}|^2d\W}{(E^0_3+E^0_4)|{\vec{\tilde{p}}}|+(E^0_3-E^0_4)|\vec p_\chi|(\vec{\tilde n}\cdot\vec n_\chi)},
\end{eqnarray}
where $d\W$ is the solid angle element of the final state momentum.
In the original frame, there is no specific three dimensional direction, and hence, the cross section is independent of the direction of $\vec P_{\rm ini}$.
We may therefore take $\vec P_{\rm ini}=(0,0,p_z)$ without loss of generality, which leads $|\vec p_\chi|=p_\chi^0|p_z|/\sqrt{S}$ and $p_\chi^0=\m_\chi\sqrt{(S+p_z^2)/S}$.
By performing the integration over the solid angle, we obtain the cross section.
At the leading order of $\m_\chi$, we have 
\begin{eqnarray}
	d\F(\tilde p_3,\tilde p_4) = d\F(p_3,p_4) + O(\m_\chi^2),~~~d\F(p_3,p_4)=\frac{d\W}{32\pi^2},
\end{eqnarray}
and hence, the phase space factors do not have $\mu_\chi$ dependence at the leading order.
Therefore, we find that the $\m_\chi$ dependence of the reaction rates, $I$,
should appear through the momentum dependence of the squared amplitude
after phase space integration,
\begin{eqnarray}
	I \simeq \int d\F(p_3,p_4)\sum_{\rm spins}|{\cal M}(p_1,p_2,\tilde p_3,\tilde p_4)|^2.
\end{eqnarray}
It should be noted that in the case where the leptons appear only in the initial states, 
$p_\chi$ dependence would appear through the initial state momenta such as ${\cal M}(\tilde p_1,\tilde p_2,p_3,p_4)$.
On the other hand, since $p_\chi$ dependence in the delta function can be always absorbed by $p_3$ and $p_4$ dependence of the amplitude.
Therefore, in this case the cross section does not depend on $\m_\chi$ since the amplitude is proportional to $(p_1\cdot p_2)$, but $(p_3\cdot p_4)$.
It is also possible that $p_\chi$ dependence is absorbed by $p_1$ and $p_2$.
However, such redefinition of initial state momenta makes the definition of ``actual" lepton chemical potential obscure, and thus, this way is rather complex\footnote{In this way we should introduce the actual lepton chemical potential so that both cases are equivalent.}, and we do not employ this manner.

\section{Cross section calculations in other field basis}
\label{sec:appendix2}

We here discuss an alternative and equivalent way to calculate cross sections by employing other field basis.
When we achieve the effective Lagrangian given by Eq.~(\ref{Leff}), $L$ and $N_R$ is rotated along to $\chi$ flat direction, and thus we have the $\del_\m\chi J^\m_L$ interaction.
On the other hand, physical observables should not be responsible to this transformation.
Therefore, we expect that our result does not change in the case that we employ the field basis without the rotation of $L$.
Let us confirm this anticipation.

The effective Lagrangian we are interested in is 
\begin{eqnarray}
	{\cal L}_{\rm eff}' &=& \text{(kinetic terms)} - \left[\frac{m_\n}{2v_{ew}^2}e^{i2\chi/(\sqrt{2}v_{B-L})}(\overline{L^C}\cdot H)(L\cdot H)+h.c.\right]+\cdots,
\end{eqnarray}
instead of ${\cal L}_{\rm eff}$ given by Eq.~(\ref{Leff}).
As mentioned in the appendix \ref{sec:appendix1}, the time-variation of $\chi$ is much slower than the time scale of the scattering process.
This allows us to expand $\chi$ by a reference time as $\chi=\dot\chi t + \cdots$, and thus we have $e^{i\chi/(\sqrt{2}v_{B-L})}\sim e^{i\m_\chi t}$ where we can $\m_\chi$ as a constant in this calculation.%
\footnote{The constant part of $\chi$ can be absorbed by the phases of the fields, and hence,
 do not have physical effects.}
A comparison between ${\cal L}_{\rm eff}$, ${\cal L}_{\rm eff}'$ leads to a replacement of the couplings
from $m_\n/(2v_{\rm ew}^2) \to m_\n/(2v_{\rm ew}^2) e^{2i\m_\chi t}$, 
without the modification of the fermion dispersion relations.
This replacement therefore modifies the delta function in ${\cal S}$ matrix, $\d^4(p_1+p_2-p_3-p_4)$, to, e.g., $\d^4(p_1+p_2-p_3-p_4 + 2p_\chi)$ as is the case of Eq.~(\ref{modified S-matrix}).
Consequently we obtain the same expression of the ${\cal S}$ matrix leading to the equivalent results of the calculation in the appendix \ref{sec:appendix1}.

\bibliographystyle{h-physrev}
\bibliography{Reference}

\end{document}